\documentclass[%
reprint,
amsmath,amssymb,
prb
]{revtex4-2}

\usepackage{footmisc}
\DefineFNsymbols*{lamport}[math]{{\dagger}{\ddagger}}
\setfnsymbol{lamport}
\usepackage{dcolumn}% Align table columns on decimal point
\usepackage{bm}% bold math

\usepackage[english]{babel}
\usepackage[utf8]{inputenc}
\usepackage{siunitx} 
\usepackage{graphicx}% Include figure files
\usepackage{caption}
\usepackage{subcaption}
\addto\captionsenglish{}
\begin{document}
\preprint{APS/123-QED}
\title{Hydrodynamic theory of the Dyakonov-Shur instability in graphene transistors}
\author{Justin Crabb}
\email{crabb.j@northeastern.edu}
\author{Xavier Cantos-Roman}
\author{Josep M. Jornet}
\affiliation{Northeastern University, Boston, Massachusetts 02115, USA}
\author{Gregory R. Aizin}
\email{GAizin@kbcc.cuny.edu}
\affiliation{Kingsborough College, The City University of New York, Brooklyn, New York 11235, USA}

\date{\today}

\begin{abstract}
We present a comprehensive theory of the Dyakonov-Shur (DS) plasma instability in current-biased graphene transistors. Using the hydrodynamic approach, we derive equations describing the DS instability in the two-dimensional electron fluid in graphene at arbitrary values of electron drift velocity. These non-linear equations together with Maxwell's equations are used for numerical analysis of the spatial and temporal evolution of the graphene electron system after the DS instability is triggered by random current fluctuations. We analyze conditions necessary for the onset of the DS instability and the properties of the final stationary state of the graphene electron system. We demonstrate that the instability results in the coherent anharmonic oscillatory state of the electron fluid and calculate both the spatial distribution and the power of the electromagnetic radiation generated by the graphene transistor in the DS instability regime.
\end{abstract}

\maketitle

\section{Introduction}

Recent years have shown a growing demand for sources of electromagnetic (EM) radiation in the THz region of the EM spectrum. This demand is driven by numerous existing and potential applications of the THz technology for security sensing and imaging systems \cite{Sensing04,federici05} as well as emerging THz communications applications \cite{Channel11,akyildiz14,mittleman17}. In the communication industry, drastically increasing data transmission rates require higher bandwidths for wireless communications \cite{Channel11}. These bandwidths are readily available by tapping into the THz band of the EM spectrum \cite{akyildiz14,mittleman17}, and THz wireless local area networks are an essential part of the next generation 6G communication systems \cite{akyildiz14teranets,polese20}. Another emerging application is THz wireless communications in nanoscale. Several nanoscale THz communications links have been proposed for intrabody communications as well as on-chip and chip-to-chip links such as wireless networks on chips \cite{israel13,moltchanov16}. Moreover, Ultra-Massive Multiple Input Multiple Output (UM MIMO) THz communication systems combat the short range of low-powered compact THz transceivers using several array modes for UM beamforming, UM spatial multiplexing, and multi-band communication schemes \cite{akyildiz16}. The above examples emphasize importance of designing tunable compact sources of THz EM radiation.

One of the promising directions in developing an on-chip tunable THz EM source is to use plasma oscillations in the two-dimensional (2D) electron channels of field-effect transistors (FETs) \cite{DS93,ryzhii05,mikhailov98,petrov17,aizin20,kachorovskii12,aizin16,knap04,lusakowski05,dyakonova05,dyakonova06,boubanga10,GaN,onishi10,otsuji13,jakvstas17}. The frequency of these oscillations lies in the THz range if the characteristic spatial scale, which determines the plasmon wave vector in the 2D channel, is of the order of \SIrange{0.1}{1}{\micro\meter}. Different physical mechanisms have been proposed to excite, and most importantly to maintain, radiating electron plasma oscillations in the transistor channel with energy supplied by an external DC electric circuit. Among them are the Dyakonov-Shur (DS) instability in asymmetric plasmonic cavities formed in the FETs \cite{DS93} and the transit time instability in the FETs with non-uniform spatial distribution of the 2D electron velocity in the transistor channel at bias voltages close to the saturation voltage \cite{ryzhii05}. Other mechanisms include reflection-type plasma instabilities \cite{mikhailov98,petrov17,aizin20} and the plasmonic boom instability \cite{kachorovskii12,aizin16}, which occur in the FETs with a grating gate or periodically changing geometry.

The DS plasma instability first predicted in \cite{DS93} occurs in the 2D electron channel of the FET under a DC current bias. Plasma waves spontaneously excited in the channel are reflected from the channel boundaries defined by the source and the drain contacts and remain confined within the plasmonic cavity formed in the channel. When a DC current passes through the transistor, the plasma waves traveling in opposite directions experience different Doppler shifts in frequency which changes after each reflection from the boundary. Dyakonov and Shur have shown that the plasma wave amplitude may increase after reflection from the boundary with a fixed total current. In this process, the energy is transferred from the DC current to the plasma wave. If the source and drain boundaries are made asymmetric, the plasma wave amplitude may increase after each round trip. This process results in the plasma instability if the plasma wave gain exceeds the damping losses. The asymmetry necessary for the DS instability is provided by the different reactive impedances between the gate and source contacts, $Z_\text{gs}$, and the gate and drain contacts, $Z_\text{gd}$. In the ideal case considered in \cite{DS93}, one should have $Z_\text{gs} = 0$ and $Z_\text{gd} = \infty$. In the final stationary state (the endpoint of instability), the power provided by the external DC circuit should be balanced by the Joule heating losses and the EM radiation emitted by the 2D electron system at the plasma frequency in the THz range.

So far, experimental efforts have mostly focused on the search for the DS instability in semiconductor FETs \cite{knap04,lusakowski05,dyakonova05,dyakonova06,boubanga10,GaN,onishi10,otsuji13,jakvstas17}. All of these works were mostly concerned with the detection of THz EM radiation expected in the final stationary state at the instability endpoint. Although weak THz radiation was recorded at bias currents exceeding some threshold value, its attribution to the DS instability was inconclusive. The radiation was mostly broadband without resonant features related to the plasmon excitations and not tunable by the gate voltage contrary to the theoretical predictions made in \cite{DS93}.

The discovery of graphene has fundamentally changed the operating limits of electronic devices with 2D electron channels \cite{neto09}. Record-high 2D electron mobility in graphene makes it possible to observe well-defined plasmon resonances with a high quality factor even at room temperature \cite{grigorenko12}, opening the door to various applications of graphene plasmonics \cite{huang17,chen17,ooi17,guo17,fan19}. In particular, recent experimental studies have demonstrated resonant detection of THz EM radiation in plasmonic cavities formed in graphene FETs \cite{bandurin18}. In another recent experiment, it was shown that the interaction of the incident THz EM radiation with the plasmons in the current-biased graphene transistor structures with a grating gate results in the amplification of the THz radiation \cite{boubanga20}. These results justify the growing interest in the DS instability in graphene transistor structures.

To date, research efforts focusing on the DS instability in graphene structures are mostly limited to theoretical studies \cite{tomadin13,svintsov13,mendl18,mendl21}. Equations describing the DS instability in graphene have been derived using the hydrodynamic model within the linear response theory \cite{tomadin13,svintsov13}. Recently, the effect of finite viscosity of the electron fluid in graphene on the DS instability was explored in the numerical model \cite{mendl18} as well as the properties of the final stationary state formed at the instability endpoint \cite{mendl21}.

In this paper, we present a comprehensive theory of the DS instability in graphene transistors, describing evolution of the instability from the very beginning (the instability threshold) till the instability endpoint, when the radiating stationary state is developed. First, we examine an analytical description of the DS instability at arbitrary allowed values of the electron drift velocity. Our approach is based on the hydrodynamic model of the electron fluid in graphene (Section~\ref{HDM}). We use the derived non-linear hydrodynamic equations together with the full system of Maxwell's equations in our original Multiphysics Simulation Platform to numerically calculate and analyze various aspects of the DS instability in graphene transistors.  This includes a numerical analysis of the conditions necessary for an onset of the DS instability, properties of the final stationary state (the instability endpoint), as well as analysis of the EM radiation emitted by the graphene transistor in the DS instability regime (Section~\ref{CompAnalysis}). Discussion of the results and concluding remarks are presented in Section~\ref{con}.

\section{Hydrodynamic Description of the 2D Electron System in Graphene} \label{HDM}
% In this section we will discuss the design of the structure and the material composition. The physical incorporation of the Dyakonov-Shur instability into the device is then explained, along with an introduction to the working principle of the entire system.  
% \subsection{HEMT Architecture}
% The architecture of the device is shown in Fig. 1. It consists of a 1 micron graphene channel sandwitched between two dielectric barriers of $SiO_2$, \SI{20}{\nano\meter} in depth on top and \SI{400}{\nano\meter} in depth on bottom.  The metal gate electrode is placed on top with a length of \SI{1}{\micro\meter} and a depth of \SI{20}{\nano\meter}.  Finally, the source and drain metal contacts are placed on the sides of the device, \SI{50}{\nano\meter} in length. The SPP wave resonates along the graphene channel. An electromagnetic wave radiates from this resonant cavity, escaping through the gaps between the source and drain contacts and the gate.  
% \subsection{Dyakonov-Shur Instability}
% The system as a whole includes the HEMT device as previously discussed as well as the external components that allow the device to resonate in the THz frequencies.  This includes a voltage source connected to the source and drain, and a current source connected from the drain to source, allowing the Dyakonov-Shur instability to take place.  |DC al Segno| 

In this section we derive the hydrodynamic equations describing the 2D electron system in graphene and apply these equations for analysis of the DS instability.

\subsection{Hydrodynamic Equations}
Hydrodynamic theory of 2D electron transport in graphene was developed in a number of publications \cite{bistritzer09,rudin11,svintsov12,tomadin13,svintsov13,briskot15}. Here, for the sake of completeness, we briefly outline main steps used to derive these equations and analyze the results.

In the quasiclassical limit, kinetic behavior of the 2D electron system in graphene is described by the electron distribution function $f_\alpha(\bm{r},\bm{p},t)$ to be found from the Boltzmann equation
\begin{equation} \label{Boltzmann}
\begin{split}
    &\frac{\partial f_\alpha(\bm{r},\bm{p},t)}{\partial t} + \bm{v}_\alpha\cdot\frac{\partial f_\alpha(\bm{r},\bm{p},t)}{\partial \bm{r}}\\
    &- e\bm{E}(\bm{r},t)\cdot\frac{\partial f_\alpha(\bm{r},\bm{p},t)}{\partial \bm{p}}= Stf_\alpha(\bm{r},\bm{p},t).
\end{split}
\end{equation}
Here, $\bm{v}_\alpha = \frac{\partial \varepsilon_\alpha(\bm{p})}{\partial \bm{p}}$ is the electron velocity, $\varepsilon_\alpha(\bm{p})$ is the electron dispersion law with index $\alpha$ referring to the band, valley, and spin quantum numbers collectively, $\bm{E}(\bm{r},t)$ is the net electric field applied to the electron with charge $-e$, and $Stf_\alpha(\bm{r},\bm{p},t)$ is the collision integral accounting for electron-electron scattering as well as electron scattering on phonons and impurities. The hydrodynamic description of electron dynamics becomes possible if the time of electron-electron collisions is much smaller than any other characteristic time in the electron system, such as electron scattering time on phonons and impurities, or inverse frequency of any external field, or electron travel time between the system’s boundaries. In this case, fast inter-electron collisions establish the local Fermi distribution function $f_{\alpha0}(\bm{r},\bm{p},t)$ characterized macroscopically by the local values of the chemical potential $\mu(\bm{r},t)$, electron temperature $T(\bm{r},t)$, and since electron-electron collisions do not change the total momentum of the interacting electrons, by the local drift velocity $\bm{v}(\bm{r},t)$. In the stationary frame of reference, the function $f_{\alpha0}(\bm{r},\bm{p},t)$ is represented by the drifting Fermi distribution function \cite{gantmakher12}
\begin{equation} \label{Fermi}
    f_{\alpha0}(\bm{r},\bm{p},t) = \frac{1}{1+e^{\frac{\varepsilon_\alpha(\bm{p})-\bm{p}\cdot\bm{v}(\bm{r},t)-\mu(\bm{r},t)}{k_\text{B}T(\bm{r},t)}}}.
\end{equation}
This function should be used in Eq.~(\ref{Boltzmann}) to find macroscopic functions $\bm{v}(\bm{r},t)$, $\mu(\bm{r},t)$, and $T(\bm{r},t)$. 

In the following, we assume that relatively small currents are driven through the graphene layer so that the released Joule heat is efficiently absorbed in the surrounding medium (the substrate) maintaining $T(\bm{r},t) = T = \text{const}$ where $T$ is the lattice temperature. We consider doped graphene layers in the degenerate limit, $\mu/T \gg 1$, so that transport occurs in the conduction band only. We also neglect inter-valley and spin scattering and omit index $\alpha$ in the following formulas.

Electron dispersion law in the graphene conduction band is $\varepsilon(\bm{p}) = v_\text{F}p$ where $v_\text{F} = \SI{1.5e6}{\meter\per\second}$ and $p = \sqrt{p_x^2+p_y^2}$ is the magnitude of the electron momentum. Using this dispersion relation and Eq.~(\ref{Fermi}) we can obtain an expression for the local electron density $n(\bm{r},t)$ in the degenerate limit
\begin{equation}\label{n}
\begin{split}
    n(\bm{r},t) &= \frac{g}{(2\pi\hbar)^2}\int f_0(\bm{r},\bm{p},t)\,d\bm{p}\\
    &= \frac{\mu^2(\bm{r},t)}{\pi\hbar^2v_\text{F}^2\left( 1-\frac{v^2(\bm{r},t)}{v_\text{F}^2}\right)^{3/2}}.
\end{split}
\end{equation}
Here, $g = 4$ is the spin and valley degeneracy factor of the 2D electrons in graphene and $\mu (\bm{r},t)$ is the local value of the chemical potential (the Fermi energy). This equation establishes the relationship between the local values of the electron density $n(\bm{r},t)$ and the Fermi energy $\mu (\bm{r},t)$ in the drifting degenerate system of massless Dirac fermions in the hydrodynamic approximation. In the frame of reference moving with the velocity $\bm{v}(\bm{r},t)$, the electron fluid is stationary, and the same electron density can be written as 
\begin{equation} \label{n2}
    n(\bm{r},t) = \frac{E_\text{F}^2(\bm{r},t)}{\pi\hbar^2v_\text{F}^2},
\end{equation}
where $E_F(\bm{r},t)$ is the local value of the Fermi energy in the stationary electron fluid. Comparing Eqs.~(\ref{n}) and (\ref{n2}) we obtain
\begin{equation}\label{mu}
    \mu = E_\text{F}\left(1 - \frac{v^2}{v_\text{F}^2}\right)^{3/4}.
\end{equation}
Equation~(\ref{mu}) determines dependence of the local Fermi energy $\mu$ on the local drift velocity $v$ in the drifting degenerate system of the massless Dirac electrons. At arbitrary temperatures, this relationship was derived in \cite{rudin11}.

In the hydrodynamic model, the average momentum $\langle \bm{p} \rangle$ and velocity $\langle \bm{v} \rangle$ per one electron can be found as
\begin{equation} \label{avg_p_v}
\begin{split}
    \langle \bm{p} \rangle &= \frac{1}{n(\bm{r},t)}\frac{g}{(2\pi\hbar)^2}\int \bm{p}f_0(\bm{r},\bm{p},t)\,d\bm{p}\\
    \langle \bm{v} \rangle &= \frac{1}{n(\bm{r},t)}\frac{g}{(2\pi\hbar)^2}\int \frac{\partial\varepsilon(\bm{p})}{\partial\bm{p}}f_0(\bm{r},\bm{p},t)\,d\bm{p}.
\end{split}
\end{equation}
After some tedious but straightforward evaluation of the integrals in Eq.~(\ref{avg_p_v}), with $f_0(\bm{r},\bm{p},t)$ defined in Eq.~(\ref{Fermi}), we obtain $\langle \bm{v} \rangle = \bm{v}(\bm{r},t)$ as expected and
\begin{equation} \label{avg_p}
    \langle \bm{p} \rangle = \frac{\mu(\bm{r},t)}{v_\text{F}^2\left(1-\frac{v^2(\bm{r},t)}{v_\text{F}^2}\right)} \langle \bm{v} \rangle =  \frac{E_\text{F}(\bm{r},t)}{v_\text{F}^2\left(1-\frac{v^2(\bm{r},t)}{v_\text{F}^2}\right)^{1/4}} \langle \bm{v} \rangle.
\end{equation}
The last equation suggests that the “hydrodynamic” effective electron mass $m_\text{H}(\bm{r},t)$ can be introduced as
\begin{equation} \label{e_mass}
    m_\text{H}(\bm{r},t) = \frac{E_\text{F}(\bm{r},t)}{v_\text{F}^2\left(1-\frac{v^2(\bm{r},t)}{v_\text{F}^2}\right)^{1/4}}.
\end{equation}
The concept of the “hydrodynamic” effective mass in the context of hydrodynamic description of the electron transport in graphene at arbitrary values of the drift velocity $v$ was introduced in \cite{svintsov13}. Our expression for $m_\text{H}(\bm{r},t)$ in Eq.~(\ref{e_mass}) is different from that derived in \cite{svintsov13}. The reason of discrepancy lies in different assumptions made in both works. In \cite{svintsov13}, the authors assumed that the local value of the chemical potential $\mu(\bm{r},t)$ in Eq.~(\ref{Fermi}) is the same in the laboratory frame of reference and in the frame of reference moving with the drift velocity $\bm{v}(\bm{r},t)$. This assumption leads to the physically controversial result that the local electron density $n(\bm{r},t)$ depends on the frame of reference.

The hydrodynamic equations (equation of continuity and the Euler equation) can be obtained as the first two moments of the Boltzmann equation~(\ref{Boltzmann}) with the electron distribution function defined in Eq.~(\ref{Fermi}) \cite{bistritzer09,rudin11,svintsov12,tomadin13,svintsov13,briskot15}. Integrating Eq.~(\ref{Boltzmann}) in the momentum space and taking into account conservation of the total number of electrons in the collisions included into $Stf(\bm{r},\bm{p},t)$, we obtain equation of continuity
\begin{equation} \label{eq7}
    \frac{\partial n(\bm{r},t)}{\partial t} + \frac{\partial}{\partial \bm{r}}\cdot[n(\bm{r},t) \bm{v}(\bm{r},t)] = 0.
\end{equation}
Multiplication of Eq.~(\ref{Boltzmann}) by the momentum $\bm{p}$ with subsequent integration in the momentum space yields the following equation
\begin{equation}\label{eq8}
\begin{split}
    &\frac{\partial}{\partial t}[n(\bm{r},t)\langle p_i(\bm{r},t)\rangle] + \frac{\partial\Pi_{ij}(\bm{r},t)}{\partial r_j} + eE_i(\bm{r},t)n(\bm{r},t) \\
    &= A_i(\bm{r},t),\quad i=x,y,
\end{split}
\end{equation}
where
\begin{equation} \label{Pi}
    \Pi_{ij}(\bm{r},t) = \frac{g}{(2\pi\hbar)^2}\int p_i \frac{\partial\varepsilon(\bm{p})}{\partial p_j}f_0(\bm{r},\bm{p},t)\,d\bm{p}
\end{equation}
and
\begin{equation} \label{A}
    A_i(\bm{r},t) = \frac{g}{(2\pi\hbar)^2}\int p_i St_\text{e-i,ph} f_0(\bm{r},\bm{p},t)\,d\bm{p}.
\end{equation}
Collision integral $St_\text{e-i,ph}f_0(\bm{r},\bm{p},t)$ in Eq.~(\ref{A}) includes electron scattering on impurities and phonons only because the total electron momentum is conserved in the electron-electron collisions.

In the following, we assume that the graphene layer is positioned in the plane $z = 0$ and $\bm{v},\bm{E}(\bm{r},t) || \mathbf{\hat{x}}$ so that all functions in the hydrodynamic equations depend on $x$-coordinates only. In this approximation, integration in Eq.~(\ref{Pi}) yields
\begin{equation}\label{Pixx}
    \Pi_{xx}(x,t) = \frac{\left(1-\frac{2v^2(x,t)}{v_\text{F}^2}\right)E_\text{F}^3(x,t)}{3\pi\hbar^2 v_\text{F}^2\left(1-\frac{v^2(x,t)}{v_\text{F}^2}\right)^{1/4}}.
\end{equation}
Substituting Eqs.~(\ref{n}), (\ref{avg_p}), and (\ref{Pixx}) into Eq.~(\ref{eq8}) we obtain the Euler equation 
\noindent\hfill
\begin{equation} \label{Euler1}
\begin{split}
    &\sqrt{\pi}\hbar\frac{\partial}{\partial t}\left[\frac{\beta n^{3/2}}{\left(1-\beta^2\right)^{1/4}}\right] \\
    &+ \frac{\sqrt{\pi}\hbar v_\text{F}}{3}\frac{\partial}{\partial x}\left[\frac{\left(1+2\beta^2\right)n^{3/2}}{\left(1-\beta^2\right)^{1/4}}\right]+eE_xn = A_x,
\end{split}
\end{equation}
\noindent
where $\beta(x,t)=v(x,t)/v_\text{F}$ is dimensionless local drift velocity. The last equation without the collision term agrees with the Euler equation for an ideal liquid of Dirac fermions derived in \cite{briskot15}. In the limit $\beta \ll 1$, Eq.~(\ref{Euler1}) reduces to the linear version of the Euler equation in \cite{svintsov12}.

To evaluate the collision term in the right-hand side of Eq.~(\ref{Euler1}), we have restricted ourselves to a simple case of elastic electron scattering on Coulomb impurities of charge $Q$ randomly distributed in the 2D plane with average density $n_i$. In this case, collision integral $St_\text{e-i}f_0(\bm{r},\bm{p},t)$ is evaluated as \cite{bistritzer09}
\begin{equation} \label{eq13}
    St_\text{e-i}f_0(\bm{r},\bm{p},t) = -\frac{f_0(\bm{r},\bm{p},t)}{\tau_p}.
\end{equation}
Here, $\tau_p$ is the transport momentum relaxation time of electrons with momentum $p$ \cite{neto09}:
\begin{equation} \label{eq14}
    \tau_p = \frac{\hbar\varepsilon(p)}{u_0^2},
\end{equation}
where $u_0^2 = n_i\left(\frac{eQ}{4\epsilon\epsilon_0}\right)^2$and $\epsilon$ is the dielectric constant of the surrounding medium. Equations~(\ref{eq13}) and (\ref{eq14}) should be used in Eq.~(\ref{A}) to find the collision term $A_x(x,t)$. Evaluating the integral in Eq.~(\ref{A}) and using Eq.~(\ref{n}) we obtain
\begin{equation} \label{Ax}
    A_x(x,t) = -\frac{\sqrt{\pi}\hbar\sqrt{n_0}\beta(x,t) n(x,t)}{\tau},
\end{equation}
where $n_0$ is an equilibrium electron density in the 2D graphene system and $\tau = \hbar E_\text{F}/u_0^2$ is the transport momentum relaxation time at the Fermi level $E_\text{F}$ in equilibrium. In the particular case of a uniform electron system in the stationary state $v(x,t) = v_0$, $n(x,t) = n_0$, $E_x(x,t) = E_0$, Eq.~(\ref{Euler1}) with the collision term given by Eq.~(\ref{Ax}) reduces to the familiar Drude-like expression
\begin{equation} \label{v0}
    v_0 = -\frac{e\tau v_\text{F}^2}{E_\text{F}}E_0,
\end{equation}
relating constant drift velocity $v_0$ and applied constant electric field $E_0$ \cite{neto09}. The total electric field in Eq.~(\ref{Euler1}) can be split into two terms $E_x(x,t) = E_0 + E_x^\text{ind}(x,t)$, where $E_x^\text{ind}(x,t)$ is the electric field induced by the fluctuations of the electron density. After some algebra, Eqs.~(\ref{Fermi}), (\ref{Ax}), and (\ref{v0}) yield
\noindent\hfill
\begin{equation} \label{Euler}
\begin{split}
    &\frac{2-\beta^2}{2\left(1-\beta^2\right)}\frac{\partial v}{\partial t} + \frac{v_\text{F}^2\left(1-\beta^2\right)}{2n}\frac{\partial n}{\partial x} + \frac{\beta^2}{2\left(1-\beta^2\right)}v\frac{\partial v}{\partial x}\\
    &+\frac{v_\text{F}\left(1-\beta^2\right)^{1/4}}{\sqrt{\pi}\hbar\sqrt{n}}eE_x^\text{ind} + \frac{\left(v-v_0\right)\left(1-\beta^2\right)^{1/4}}{\tau}\sqrt{\frac{n_0}{n}} = 0.
\end{split}
\end{equation}
The Euler equation~(\ref{Euler}) differs from the similar equation derived in \cite{svintsov13} in the collisionless limit by the effective “hydrodynamic” mass in the field term as discussed earlier in this Section. Hydrodynamic equations~(\ref{eq7}) and (\ref{Euler}) will be used in our numerical studies of the plasma oscillations in the driven 2D electron gas in graphene in Section~\ref{CompAnalysis}.

\subsection{The DS Instability in the Graphene Transistor} \label{ds inst in graphene}
Plasma waves confined in the current-biased 2D transistor channel are subject to the DS instability if asymmetric boundary conditions are imposed at the opposite ends of the plasmonic cavity formed in the channel \cite{DS93}. To find conditions necessary for development of the DS instability in the graphene layers, we first obtain the spectra of the plasma waves in the presence of the steady electron drift with velocity $v_0$. Following the standard procedure \cite{tomadin13,svintsov13} we linearize hydrodynamic equations~(\ref{eq7}) and (\ref{Euler}) with respect to the small fluctuations of the electron density $\delta n$ and drift velocity $\delta v$:
\begin{equation} \label{eq18}
\begin{split}
    n(x,t) &= n_0 + \delta ne^{\text{i}qx-\text{i}\omega t},\\
    v(x,t) &= v_0 + \delta ve^{\text{i}qx-\text{i}\omega t}.
\end{split}
\end{equation}

Self-consistent electric field in the Euler equation~(\ref{Euler}) is determined as $E_x = -\frac{\partial\delta\phi}{\partial x}$ where $\delta\phi$ is the electric potential induced by the charge perturbation $-e\delta n$. In the quasi-static limit of the gated 2D electron channel, the values of $\delta\phi$ and $\delta n$ are related as \cite{DS93}:
\begin{equation} \label{eq19}
    \delta \phi_{q\omega} = -\frac{ed}{\epsilon\epsilon_0}\delta n_{q\omega},
\end{equation}
where $d$ is the distance between the graphene layer and the gate. This local approximation is valid for the long wavelength fluctuations of the electron density when $qd \ll 1$. Using Eqs.~(\ref{eq18}) and (\ref{eq19}) and neglecting the collision term in the Euler equation justified at $\omega\tau \gg 1$, we obtain the system of linear equations for the fluctuations of the electric potential $\delta\phi_{q\omega}$ and the current density $\delta j_{q\omega} = -e(n_0 \delta v_{q\omega} + v_0 \delta n_{q\omega}$):
\begin{equation} \label{eq20}
    q\delta j_{q\omega} - \frac{\epsilon\epsilon_0}{d}\omega\delta\phi_{q\omega} = 0,
\end{equation}
\begin{equation} \label{eq21}
\begin{split}
        &\left[\gamma\left(\omega-qv_0\right)+qv_0\right]\left(\omega-qv_0\right)\delta j_{q\omega}\\
    &- \frac{\epsilon\epsilon_0}{d}\left[\frac{v_\text{F}^2}{2}\left(1-\beta_0^2\right) + \frac{e^2n_0v_\text{F}^2d}{\epsilon\epsilon_0E_\text{F}}\left(1-\beta_0^2\right)^{1/4}\right]\delta\phi_{q\omega} = 0,
\end{split}
\end{equation}
where $\gamma = \frac{2-\beta_0^2}{2\left(1-\beta_0^2\right)}$ and $\beta_0 = v_0/v_\text{F}$. The system of Eqs.~(\ref{eq20}) and (\ref{eq21}) has non-trivial solution of $\omega = v_\text{p}^{(\pm)}q$, where
\begin{equation} \label{eq22}
    v_\text{p}^{(\pm)} = \left(1-\frac{1}{2\gamma}\right)v_0\pm\frac{1}{\gamma}\sqrt{\frac{v_\text{F}^2}{2}+\frac{\gamma e^2 n_0 v_\text{F}^2 \left(1-\beta_0^2\right)^{1/4} d}{\epsilon\epsilon_0 E_\text{F}}}.
\end{equation}
Here, $v_\text{p}^{(\pm)}$ are velocities of the plasma waves traveling in the direction of drift ($+$) and in the opposite direction ($-$). At small velocities, $v_0 \ll v_\text{F}$, Eq.~(\ref{eq22}) reduces to \cite{svintsov13}
\begin{equation} \label{eq23}
    v_\text{p}^{(\pm)} = \frac{v_0}{2} \pm v_\text{p},
\end{equation}
where
\begin{equation} \label{vp}
    v_\text{p} = \sqrt{\frac{v_\text{F}^2}{2} + \frac{e^2 n_0 v_\text{F}^2 d}{\epsilon\epsilon_0E_\text{F}}}
\end{equation}
is the velocity of the plasma waves in the 2D graphene layer in the absence of drift \cite{tomadin13,svintsov13}.

\begin{figure}[h]
\centering
  \begin{subfigure}{.48\textwidth}
        \centering
        \includegraphics[width=\linewidth]{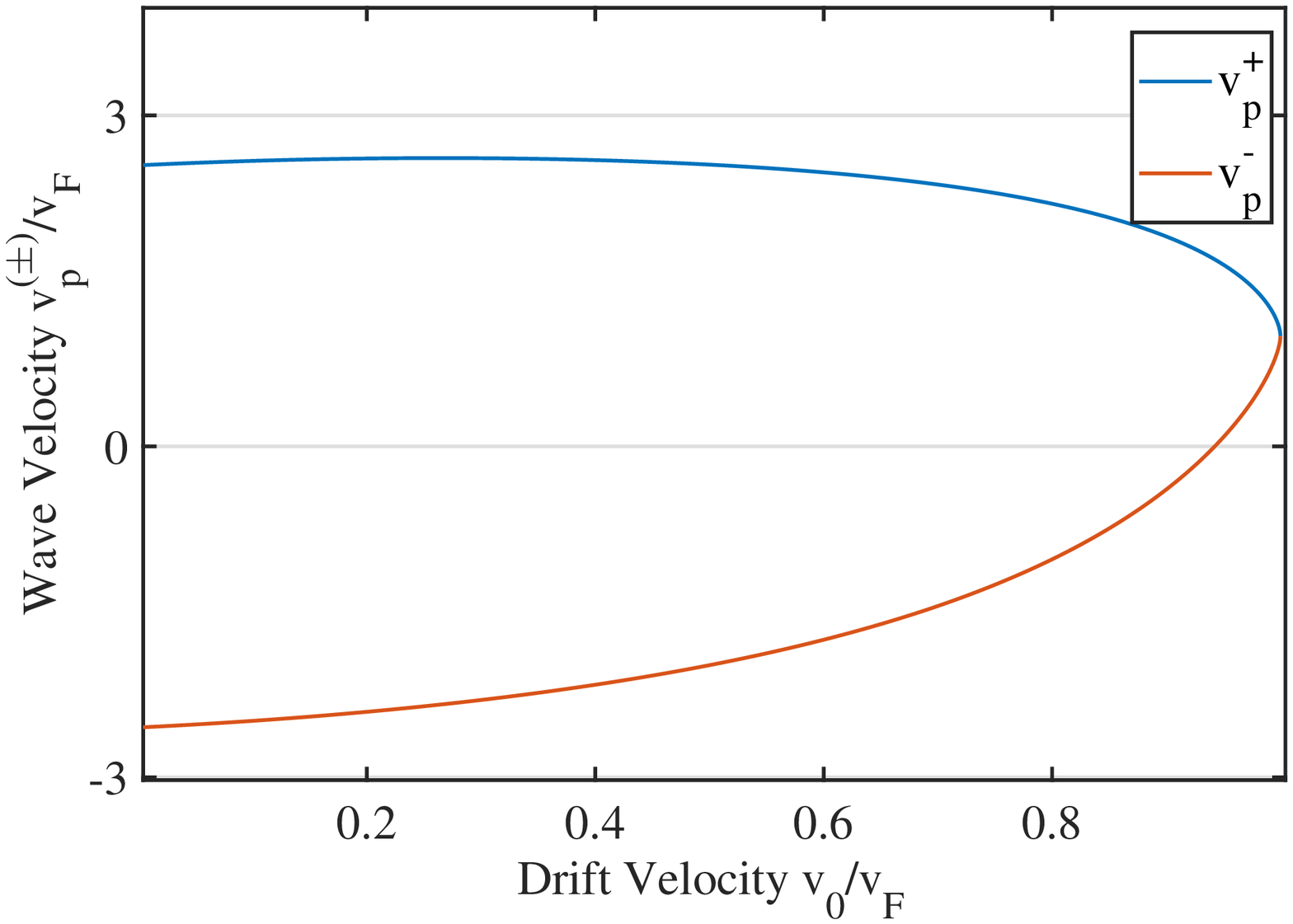}
        \caption{}
        \label{fig:s}
  \end{subfigure}
  \begin{subfigure}{.48\textwidth}
        \centering
        \includegraphics[width=\linewidth]{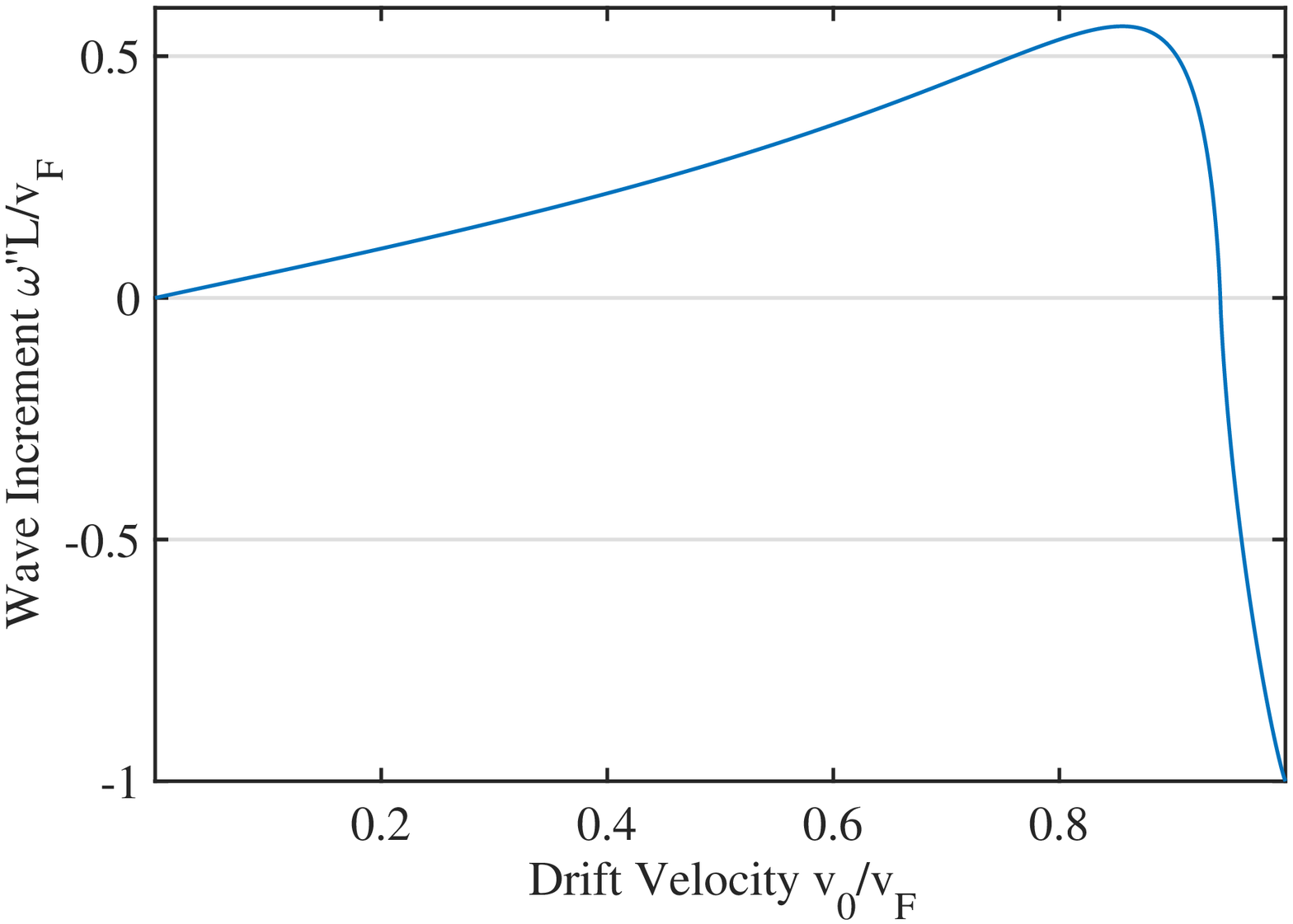}
        \caption{}
        \label{fig:gamma}
  \end{subfigure}
  \caption{(a) The wave velocities of the plasmons traveling in the direction of the DC electron drift ($v_\text{p}^{(+)}$) and in the opposite direction ($v_\text{p}^{(-)}$) as a function of the drift velocity $v_0$. (b) The plasma wave increment $\omega''$ as a function of the drift velocity $v_0$.}
  \label{fig:plasmon}
\end{figure}

Dependence of the plasma velocities $v_\text{p}^{(\pm)}$ on the drift velocity $v_0$ at arbitrary $0 \leq v_0 \leq v_\text{F}$ is shown in Fig.~\ref{fig:plasmon}(\subref{fig:s}). It follows from Eq.~(\ref{eq22}) that $v_\text{p}^{(\pm)} \rightarrow v_\text{F}$ when $v_0 \rightarrow v_\text{F}$ and does not depend on the direction of propagation of the plasma waves at $v_0 = 0$ as opposed to the result obtained in \cite{svintsov13}. Qualitatively, this result is expected because $v_\text{F}$ is an ultimate theoretical value of the drift velocity in the system of massless Dirac fermions, and the limit $v_0 \rightarrow v_\text{F}$ is equivalent to the limit $v_0 \rightarrow \infty$ in the system of electrons with finite effective mass where $v_\text{p}^{(\pm)} \rightarrow v_0$ when $v_0 \rightarrow \infty$. 

The DS instability can be obtained if Eqs.~(\ref{eq20}) and (\ref{eq21}) are complemented by the Dyakonov-Shur boundary conditions at the opposite ends of the plasmonic cavity of length $L$
\begin{equation} \label{eq25}
\begin{split}
    \delta\phi_{q\omega}(x=0) &= 0,\\
    \delta j_{q\omega}(x=L) &= 0
\end{split}
\end{equation}
corresponding to the zero gate-to-channel impedance at one end of the cavity ($x=0$) and the infinite gate-to-channel impedance at the opposite end ($x=L$) \cite{DS93}. Solving Eqs.~(\ref{eq20}) and (\ref{eq21}) with boundary conditions (\ref{eq25}), we find the complex plasma frequencies $\omega = \omega'+ \text{i}\omega''$
\begin{align} \label{eq26}
     \omega' &= \frac{v_\text{p}^{(+)}v_\text{p}^{(-)}}{v_\text{p}^{(-)}-v_\text{p}^{(+)}}\frac{\pi}{L}\begin{cases}
        2n, & v_\text{p}^{(+)}/v_\text{p}^{(-)} > 0 \\
        2n-1, & v_\text{p}^{(+)}/v_\text{p}^{(-)} < 0 \\
    \end{cases} n = 1, 2, ...\\
\label{eq27}
\omega'' &= \frac{1}{L}\frac{v_\text{p}^{(+)}v_\text{p}^{(-)}}{v_\text{p}^{(-)}-v_\text{p}^{(+)}} \ln\left|\frac{v_\text{p}^{(+)}}{v_\text{p}^{(-)}}\right|.
\end{align}

The instability arises when $\omega'' > 0$. In Fig.~\ref{fig:plasmon}(\subref{fig:gamma}) we plotted $\omega''$ as a function of $v_0$ demonstrating that the instability can occur in a broad interval of drift velocities. In the presence of collisions, the onset of the instability occurs when $\omega'' > 1/2\tau$ where $\tau$ is the electron momentum relaxation time \cite{DS93}. In the next Section we find a rigorous numerical solution of the non-linear hydrodynamic equations~(\ref{eq7}) and (\ref{Euler}) together with the full system of Maxwell’s equations replacing Eq.~(\ref{eq19}), and analyze the final stationary state of the current driven 2D electron fluid in graphene in the DS instability regime.

\section{Numerical Analysis of the DS Instability in the 2D Graphene Layer} \label{CompAnalysis}
Temporal evolution of the DS instability and potential final stationary state in the 2D electron system in graphene can be found from the numerical solution of the non-linear hydrodynamic equations~(\ref{eq7}) and (\ref{Euler}) combined with Maxwell's equations for the EM field generated by the charge density fluctuations in the 2D electron fluid. This system of equations should be solved self-consistently taking into account the boundary conditions (\ref{eq25}). So far, a self-consistent numerical solution of the hydrodynamic equations in the DS instability regime was mostly considered for semiconductor FETs with a 2D electron channel in the quasi-static limit \cite{Dmitriev97,cheremisin02,li17}. This approach does not allow direct evaluation and analysis of the THz EM radiation expected in the final stationary state of the 2D system and regarded as the most important outcome of the DS instability. The numerical solution of the DS instability problem with the full system of the Maxwell equations instead of the static Poisson equation was developed for the ungated 2D electron gas in semiconductor heterostructures in \cite{bhardwaj16} and for the III-V semiconductor-based HEMT in \cite{nafari18}. Very recently, the numerical solution of the DS instability problem in the graphene-based transistor was developed in \cite{mendl21} using the quasi-static model for the EM part of the problem. The authors analyzed the final stationary state of the 2D electron system and estimated the maximum total THz EM power generated in this state.

Below, we present the numerical solution of the DS instability problem in the graphene transistor using the full system of Maxwell's equations for the description of the generated EM field. We analyze the final stationary state of the 2D electron fluid in the transistor channel as well as the spectral content and spatial distribution of the accompanying THz EM radiation.

%The sets of equations described in the previous section are solved in discretized finite cells in a 2D grid space, providing the robustness and complexity needed to fine-tune our system and produce more realistic results.  The device configuration (shown in Fig. \ref{fig:structure}) is as follows: a graphene channel is sandwiched between two dielectric slabs of $\text{SiO}_{\text{2}}$, which lie between the metallic source and drain contacts.  A metallic gate is placed above the top dielectric slab which spans across the length of the channel. A voltage source $V_{gs}$ is attached from gate to source to induce the short circuit boundary condition, and a current source $I_{ds}$ is attached from drain to source to induce the open circuit boundary condition.  

\begin{figure}[t!]
  \includegraphics[width=\linewidth]{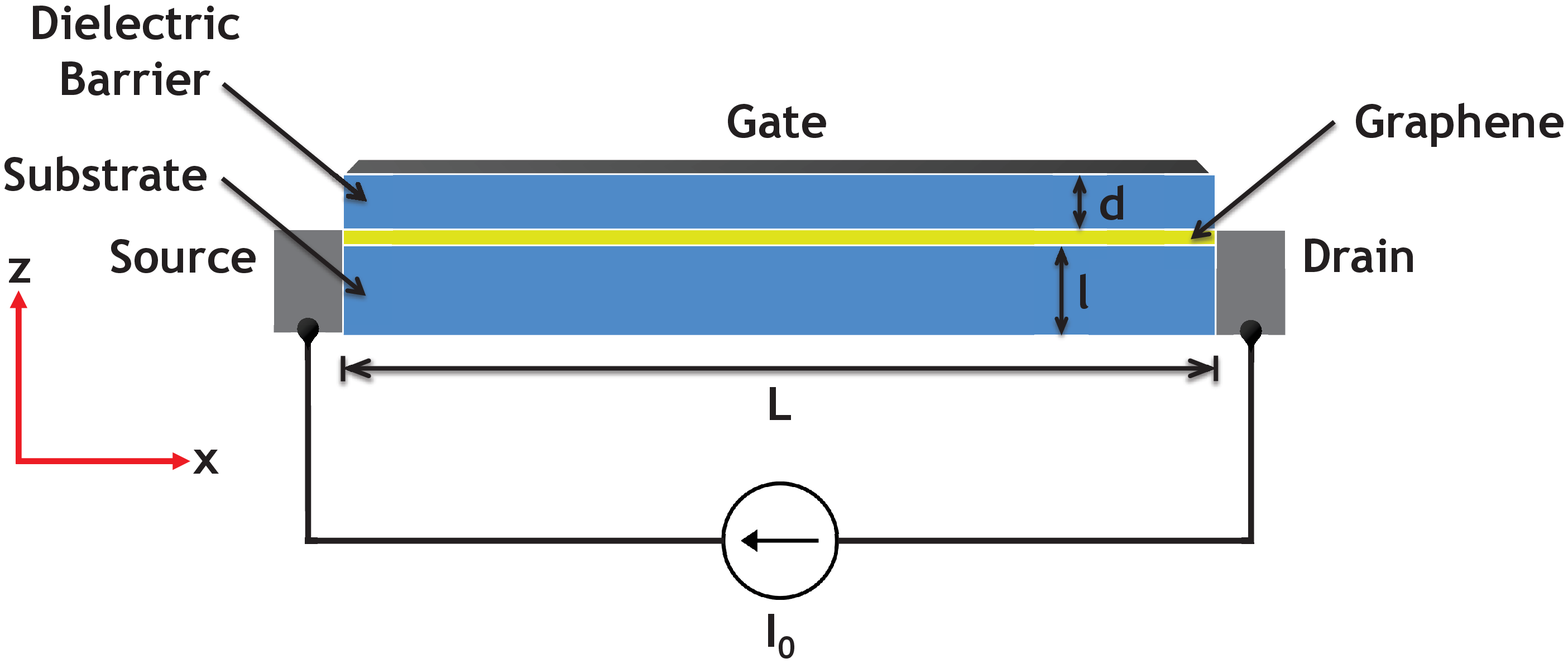}
  \caption{Schematic of the graphene transistor structure used in the numerical simulations of the DS instability.}
  \label{fig:structure}
\end{figure}

\subsection{Numerical Model}
In Fig.~\ref{fig:structure}, we show the schematic of the graphene-based transistor structure used in our numerical simulation. It consists of the graphene layer of length $L$ placed between two dielectric slabs with relative permittivity $\epsilon$, representing the substrate of thickness $l$ and the barrier layer of thickness $d$ separating the graphene channel with the 2D electron gas and the metal gate. The source and drain metal contacts are used to provide constant bias current characterized by the particle current density $j_0 = n_0v_0$ where $n_0$ is the equilibrium 2D electron density in the graphene channel and $v_0$ is the drift velocity determined by the applied constant source-drain voltage. We assume that our system is uniform in the direction perpendicular to the current in the 2D plane so that the 2D electron fluid in the graphene layer is described by the one-dimensional hydrodynamic equations~(\ref{eq7}) and (\ref{Euler}). These non-linear equations written in terms of the electron density $n(x,t)$ and the particle current density $j(x,t)=n(x,t)v(x,t)$, subject to the boundary conditions $n(x=0,t)=n_0$ and $j(x=L,t)=j_0$, should be solved numerically together with Maxwell's equations
\begin{equation}
\begin{split}
    \nabla \times \bm{E} &= -\mu_0 \frac{\partial\bm{H}}{\partial t},\\
    \nabla \times \bm{H} &= -e(j-j_0)\delta(z)\mathbf{\hat{x}} +\epsilon\epsilon_0 \frac{\partial\bm{E}}{\partial t},\\
\end{split}
\end{equation}
where $\bm{E} = E_x\mathbf{\hat{x}} + E_z\mathbf{\hat{z}}$ and $\bm{H} = H_y\mathbf{\hat{y}}$ are electric and magnetic components of the EM field induced by the fluctuations of the electric current in the channel $-e(j-j_0)$. The numerical simulation was performed using the original Finite-Difference-Time-Domain (FDTD) Multiphysics Simulation Platform because commercial tools do not allow simultaneous simulation of both hydrodynamic and Maxwell's equations in the time-domain. The platform allows both solvers to run simultaneously on MATLAB and provides a self-consistent solution of the equations discretized in the 2D grid space \cite{taflove05}. The discretization procedure and the FDTD method are described in more detail in \cite{nafari18}. In the numerical 2D grid scheme used in this simulation, the graphene channel was represented as an infinitely thin sheet lying at the boundary between two grid cells. This can be done with special choice of the grid scheme which overlays the current density $j$ and electric field $E_x$ between two separate solvers for the hydrodynamic and electrodynamic equations. The numerical algorithm also includes additional boundary conditions $\frac{\partial j(x=0,t)}{\partial x} = 0$ and $\frac{\partial n(x=L,t)}{\partial x} = 0$ at the boundaries between the 2D electron channel and the source/drain contacts. These conditions follow from the hydrodynamic equations and provide necessary matching between the graphene channel and the metal contacts. 

In our numerical simulations of the device shown in Fig.~\ref{fig:structure} we take the length of the graphene channel $L = \SI{1}{\micro\meter}$, the gate-to-channel distance $d = \SI{20}{\nano\meter}$, the substrate thickness $l = \SI{400}{\nano\meter}$, relative permittivity of the dielectric slabs $\epsilon = 3.8$, and the equilibrium 2D electron density $n_0 = \SI[per-mode=reciprocal]{4.2e16}{\per\meter\squared}$. All metal contacts were assigned an infinite conductivity to save computational time. 

\subsection{Results}
In the following sections we analyze the results of the numerical simulations of the DS instability in the graphene transistor obtained within our Simulation Platform. 

\begin{figure*}[t]
\centering
  \begin{subfigure}{.48\textwidth}
        \centering
        \includegraphics[width=.95\linewidth]{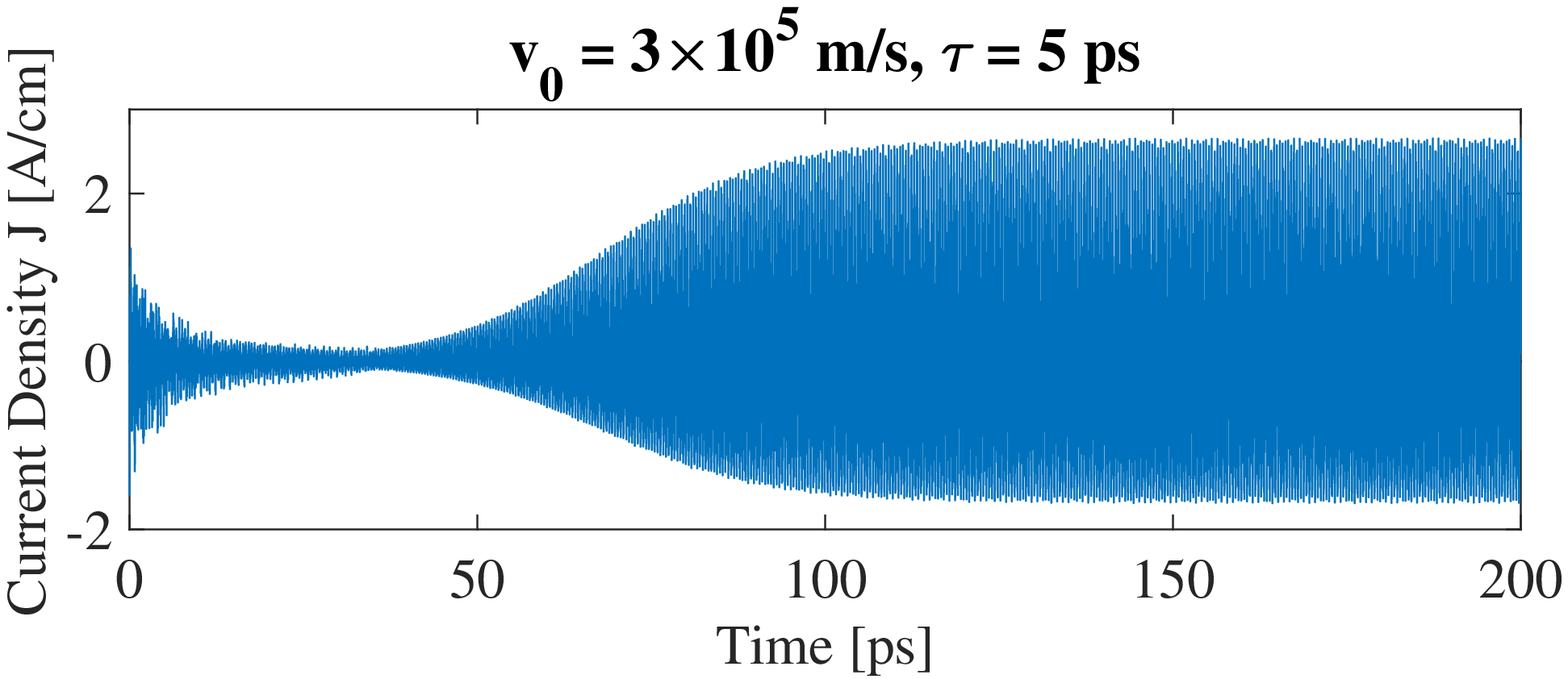}
        \caption{}
        \label{fig:v0-high}
  \end{subfigure}
  \begin{subfigure}{.48\textwidth}
        \centering
        \includegraphics[width=.95\linewidth]{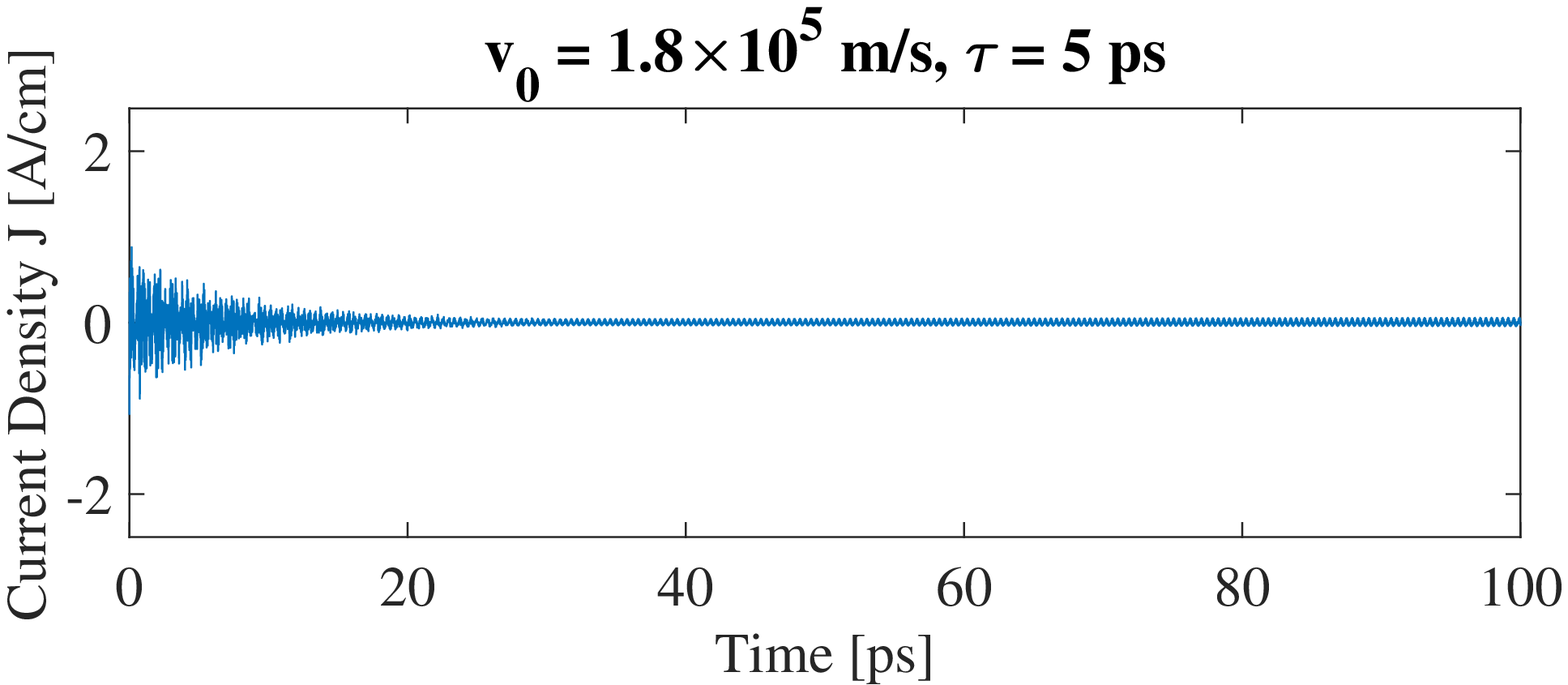}
        \caption{}
        \label{fig:v0-low}
  \end{subfigure}
  \begin{subfigure}{.48\textwidth}
        \centering
        \includegraphics[width=.95\linewidth]{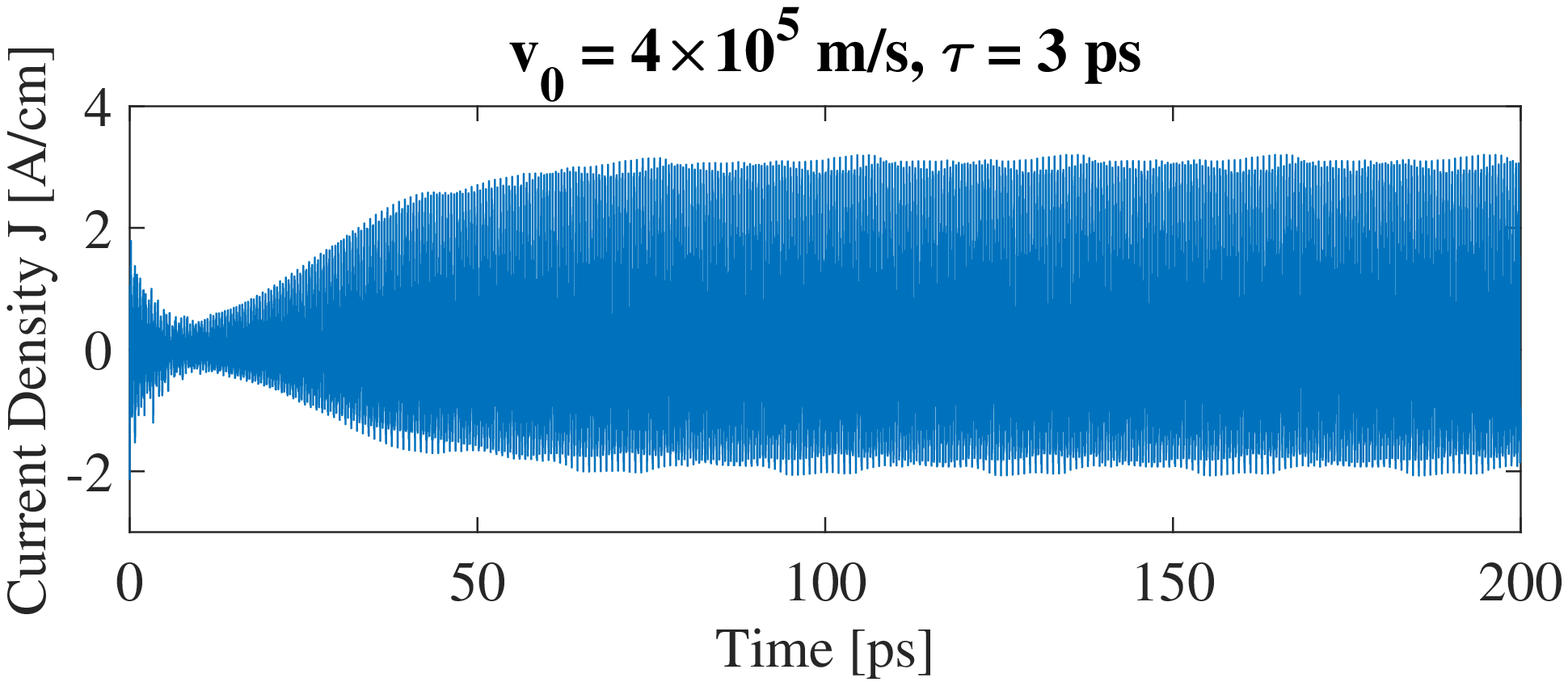}
        \caption{}
        \label{fig:tau-high}
  \end{subfigure}
  \begin{subfigure}{.48\textwidth}
        \centering
        \includegraphics[width=.95\linewidth]{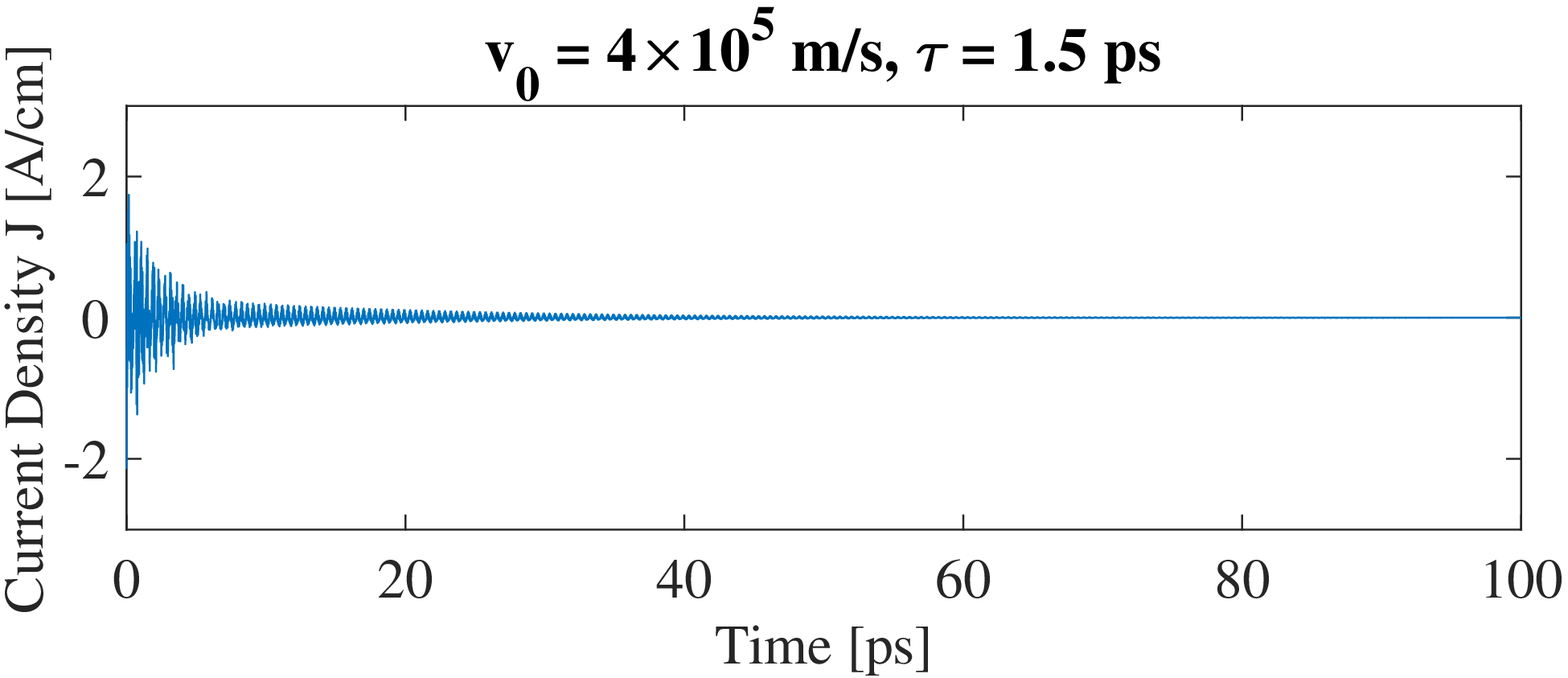}
        \caption{}
        \label{fig:tau-low}
  \end{subfigure}
  \caption{Temporal evolution of the plasmonic current density in the graphene transistor channel above ((a) and (c)) and below ((b) and (d)) the instability threshold $v_0 \geq \frac{L}{\tau}$ . The channel length $L = \SI{1}{\micro\meter}$ in all plots.}
  \label{fig:lower}
\end{figure*}

\subsubsection{The Instability Threshold}
To initiate the transistor structure response, we introduced the fluctuation of the particle current density at the initial moment of time (the ''kick''). The fluctuations with a magnitude equal to 8\% of the equilibrium particle current density $j_0$ were placed at several random cells of the numerical grid in the graphene channel. We note that the fluctuation magnitude and position do not affect the final stationary state of the system but may affect the time it takes to set up the collective plasma oscillation in the channel at the very beginning of the process. 

In Fig.~\ref{fig:lower} we show temporal evolution of the plasmonic current after the initial excitation. The current was recorded in the midpoint of the graphene channel at several different values of the drift velocity $v_0$ and relaxation time $\tau$. The plots in Figs.~\ref{fig:lower}(\subref{fig:v0-high}) through \ref{fig:lower}(\subref{fig:tau-low}) demonstrate that after some transient time the collective plasma oscillations in the 2D electron channel either exponentially decay as shown in Figs.~\ref{fig:lower}(\subref{fig:v0-low}) and \ref{fig:lower}(\subref{fig:tau-low}) or develop the instability as shown on Figs.~\ref{fig:lower}(\subref{fig:v0-high}) and \ref{fig:lower}(\subref{fig:tau-high}). The instability develops if $\omega'' > 1/2\tau$ where $\omega''$ is determined by Eq.~(\ref{eq27}). At $v_0 \ll v_\text{F}$, Eqs.~(\ref{eq23}) and (\ref{eq27}) yield the instability threshold $v_0^\text{th} > L/\tau$. This conclusion is quantitatively confirmed in our numerical simulations shown in Fig.~\ref{fig:lower}. In Figs.~\ref{fig:lower}(\subref{fig:v0-high}) and \ref{fig:lower}(\subref{fig:v0-low}) the value of relaxation time $\tau = \SI{5}{\pico\second}$ was used, with a threshold drift velocity $v_0^\text{th} = \SI{2e5}{\meter\per\second}$. In Fig.~\ref{fig:lower}(\subref{fig:v0-high}), we have $v_0 = \SI{3e5}{\meter\per\second}$ so that $v_0 > v_0^\text{th}$, whereas in Fig.~\ref{fig:lower}(\subref{fig:v0-low}), $v_0 = \SI{1.8e5}{\meter\per\second}$ and $v_0 < v_0^\text{th}$. The instability develops in Fig.~\ref{fig:lower}(\subref{fig:v0-high}) and plasma oscillations exponentially decay in Fig.~\ref{fig:lower}(\subref{fig:v0-low}). Similar behavior is observed in Figs.~\ref{fig:lower}(\subref{fig:tau-high}) and \ref{fig:lower}(\subref{fig:tau-low}) where the drift velocity is kept constant at $v_0 = \SI{4e5}{\meter\per\second}$ but simulations are performed at two different relaxation times: $\tau = \SI{3}{\pico\second}$ in Fig.~\ref{fig:lower}(\subref{fig:tau-high}) and $\tau = \SI{1.5}{\pico\second}$ in Fig.~\ref{fig:lower}(\subref{fig:tau-low}). The threshold value of the relaxation time is $\tau^\text{th} = L/v_0 = \SI{2.5}{\pico\second}$. The instability develops at $\tau > \tau^\text{th}$ in Fig.~\ref{fig:lower}(\subref{fig:tau-high}), and oscillations decay at $\tau < \tau^\text{th}$ in Fig.~\ref{fig:lower}(\subref{fig:tau-low}). 

\subsubsection{The Final Stationary State (Instability Endpoint)}

\begin{figure*}[t]
    \centering
    \begin{subfigure}[t]{.48\textwidth}
        \centering\includegraphics[width=.95\linewidth]{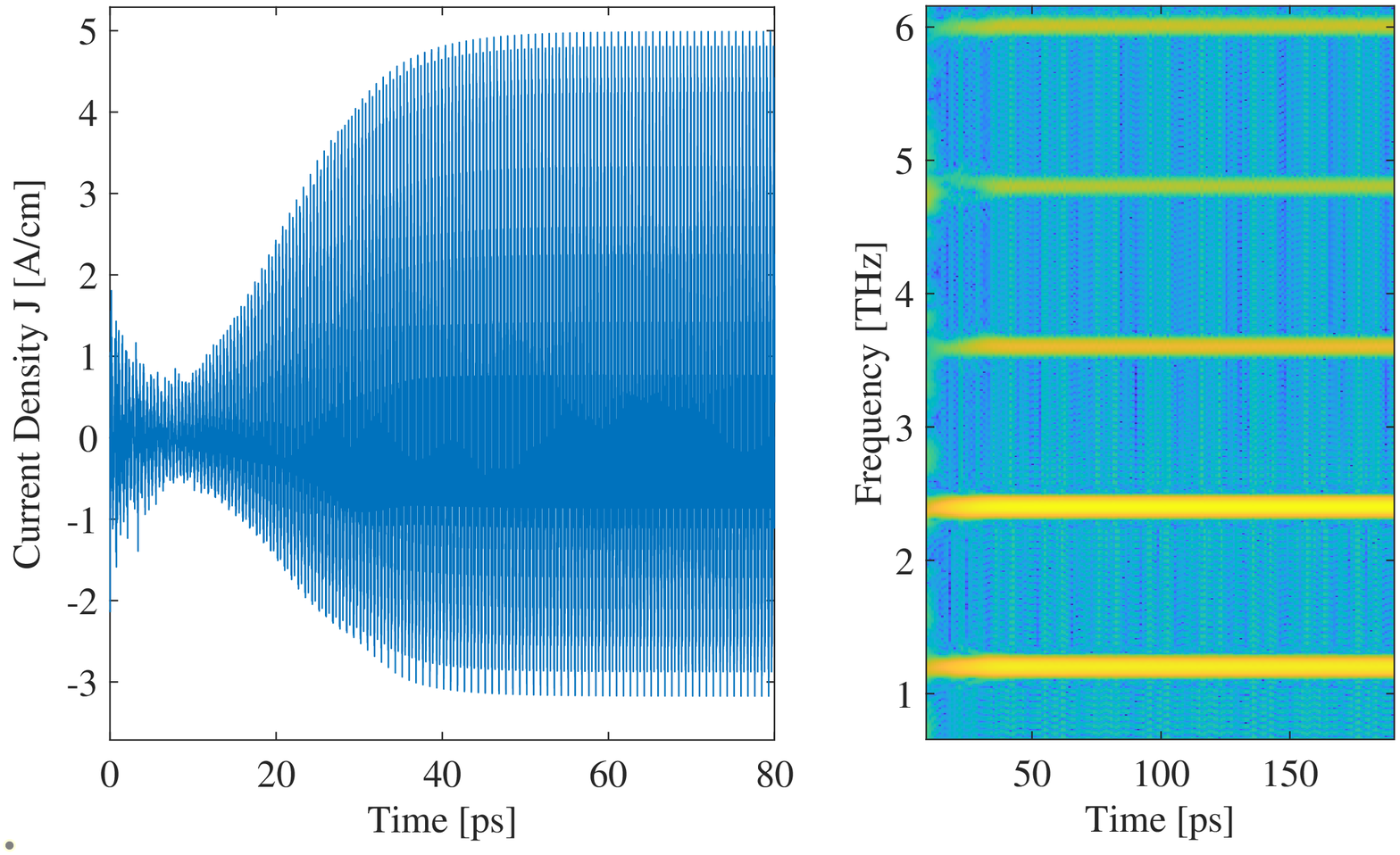}
        \caption{} 
        \label{fig:sig}
    \end{subfigure}
    %\hfill
    \begin{subfigure}[t]{.48\textwidth}
        \centering\includegraphics[width=.95\linewidth]{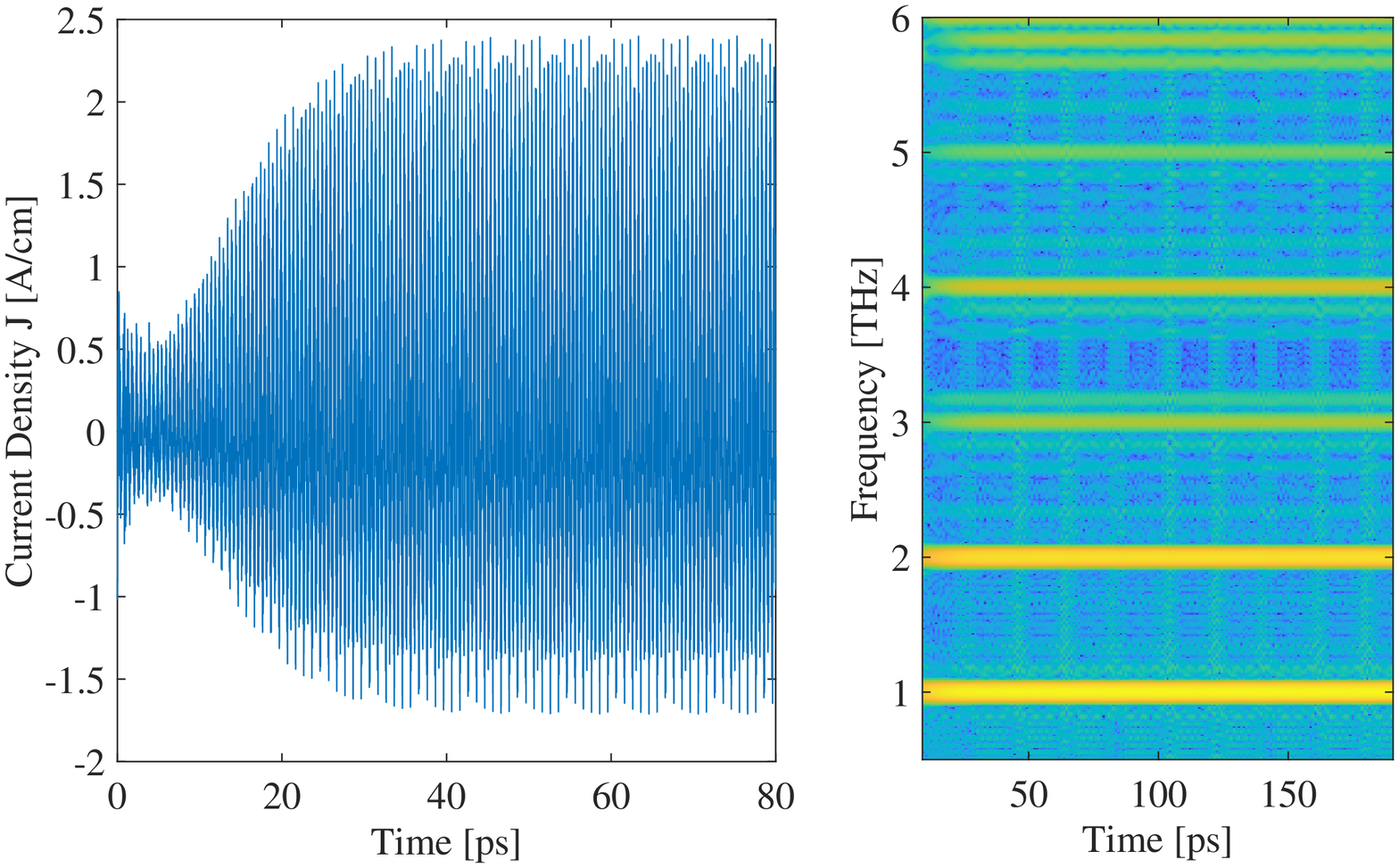}
        \caption{} 
        \label{fig:sig_n0-2}
    \end{subfigure}
    %\hfill
    \begin{subfigure}[t]{.48\textwidth}
        \centering\includegraphics[width=.95\linewidth]{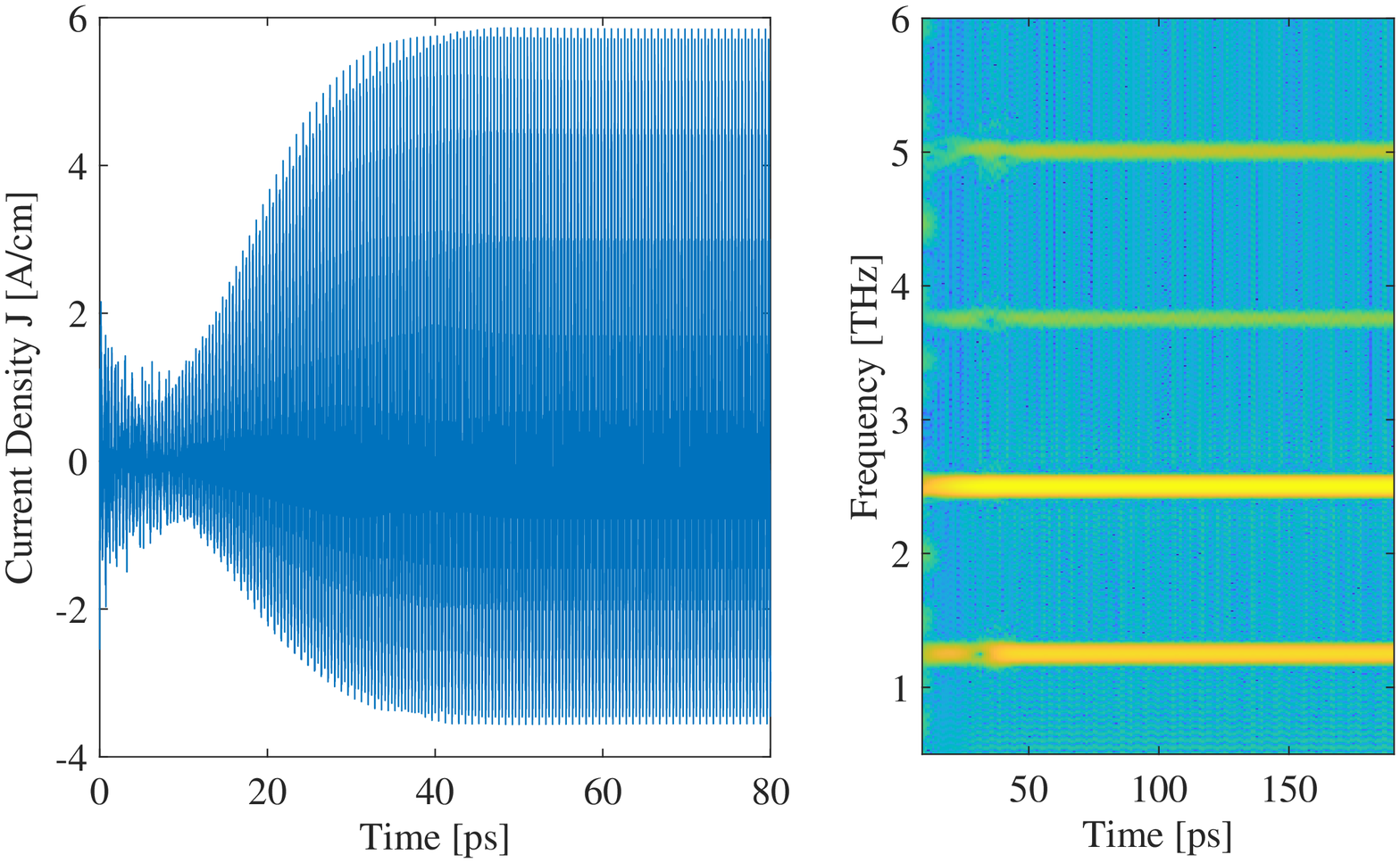} %[width=\linewidth]
        \caption{} 
        \label{fig:sig_n0-5}
    \end{subfigure}
    %\hfill
    \begin{subfigure}[t]{.48\textwidth}
        \centering\includegraphics[width=.95\linewidth]{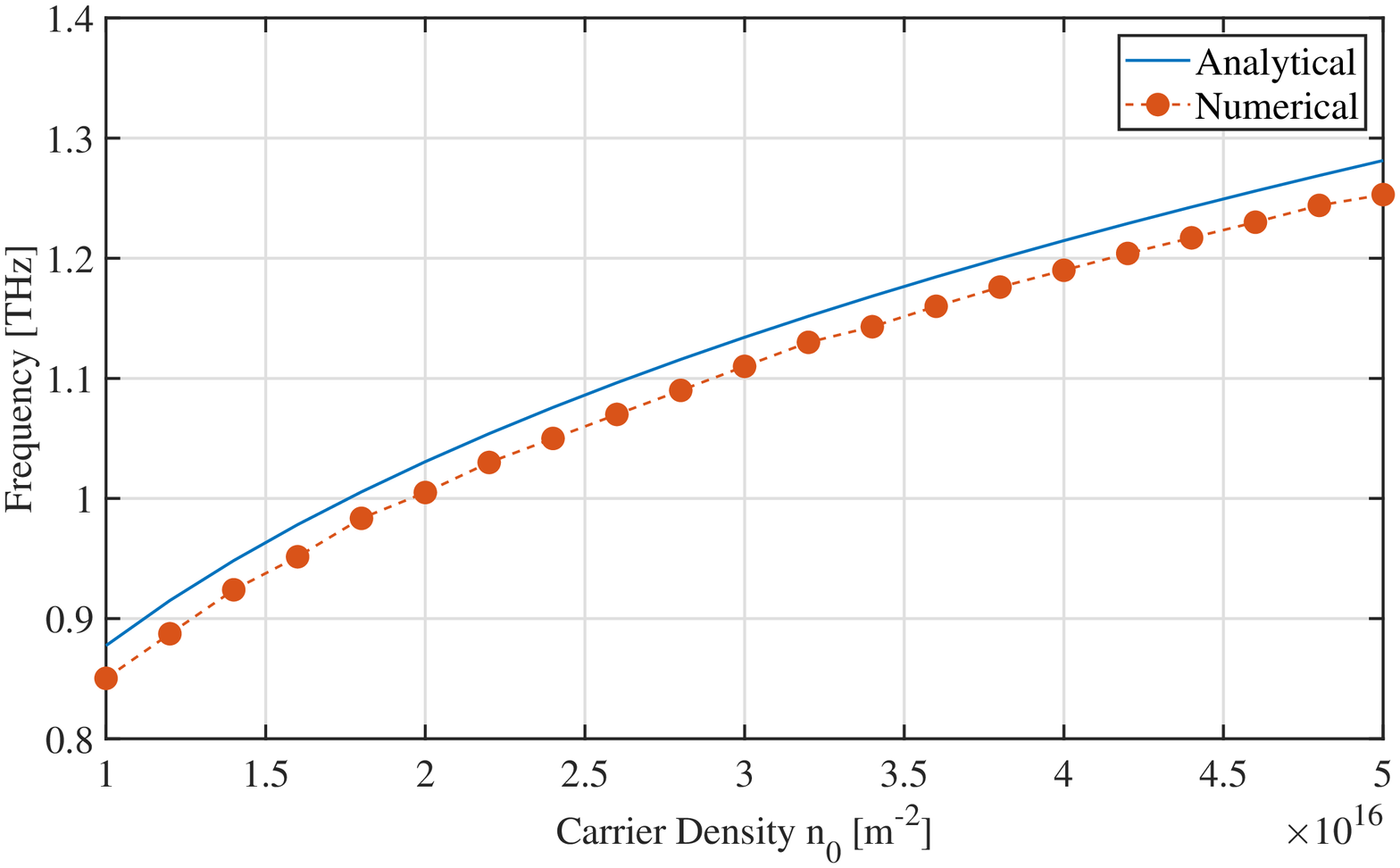} %[width=\linewidth]
        \caption{} 
        \label{fig:sig_n0}
    \end{subfigure}
    \caption{(a)-(c) The plasmonic current density in the graphene transistor channel (left) and the current spectral content (right) in the final stationary state at different electron densities: (a) $n_0 = \SI[per-mode=reciprocal]{4e16}{\per\meter\squared}$; (b) $n_0 = \SI[per-mode=reciprocal]{2e16}{\per\meter\squared}$; (c) $n_0 = \SI[per-mode=reciprocal]{5e16}{\per\meter\squared}$. In all plots $v_0 = \SI{4e5}{\meter\per\second}$, $\tau = \SI{5}{\pico\second}$. (d) The fundamental frequency of the plasmonic current oscillations as a function of electron density $n_0$ found analytically from Eq.~\ref{f0} (solid blue line) and numerically (dotted orange line).} 
    \label{fig:signal}
\end{figure*}

The central question of the DS instability problem is the character of the final state after the electron system is stabilized in the dynamic equilibrium when the energy supplied by the external DC circuit is balanced by the losses due to Joule heating and EM radiation. In semiconductor structures this problem was considered in \cite{bhardwaj16,nafari18}. In graphene, the problem of the instability endpoint was first addressed in the very recent paper by C. B. Mendl \textit{et al.} \cite{mendl21}. In all these studies it was concluded that the instability endpoint represents some coherent (non-chaotic) non-linear oscillator with electron density and plasmonic current periodically changing in time. We present the results of our studies of the final stationary state in the DS instability regime in Fig.~\ref{fig:signal}. In Figs.~\ref{fig:signal}(\subref{fig:sig}) through \ref{fig:signal}(\subref{fig:sig_n0-5}), we show the temporal evolution of the plasmonic current after an initial kick described earlier in the text at three different values of the equilibrium electron density $n_0$. These plots demonstrate that after an initial rise due to the DS instability, the signal stabilizes in some stationary periodic pattern. The spectral content of this pattern in the time domain is presented in the spectrograms placed next to each signal plot in Figs.~\ref{fig:signal}(\subref{fig:sig})-\ref{fig:signal}(\subref{fig:sig_n0-5}). These spectrograms demonstrate stable spectra in the final stationary state of the electron system in the graphene channel. The spectrum consists of a series of peaks at integer multiples of some fundamental frequency $f_0$. The peak amplitude gradually decreases for higher harmonics. This type of spectrum describes the anharmonic plasmonic current oscillations: $j(t+T) = j(t)$ with period $T = 1/f_0$ and corresponds to the coherent non-linear oscillator in agreement with previous studies \cite{bhardwaj16,nafari18,mendl21}. The fundamental frequency $f_0$ depends on $n_0$ and with high degree of accuracy coincides with the frequency of the fundamental mode found in the linear analysis of the DS instability and given by Eq.~(\ref{eq26}). In Fig.~\ref{fig:signal}(\subref{fig:sig_n0}), we plot the value of $f_0$ found in our numerical simulations at different electron densities $n_0$ and the analytical results from Eq.~(\ref{eq26}) at $n = 1$ confirming this conclusion. Numerical simulations performed at different drift velocities $v_0$ show that the value of $f_0$ does not depend on $v_0$ as long as $v_0 \ll v_\text{F}$ and are described by the simple formula
\begin{equation} \label{f0}
    f_0 = \frac{v_\text{p}}{4L},
\end{equation}
corresponding to the quarter-wavelength standing plasmonic wave in the linear response theory considered in Section~\ref{ds inst in graphene}. 

\subsubsection{The Generated Electromagnetic Fields}
\begin{figure}[ht]
    %\centering
    \begin{subfigure}{.48\textwidth}
        \includegraphics[width=\linewidth]{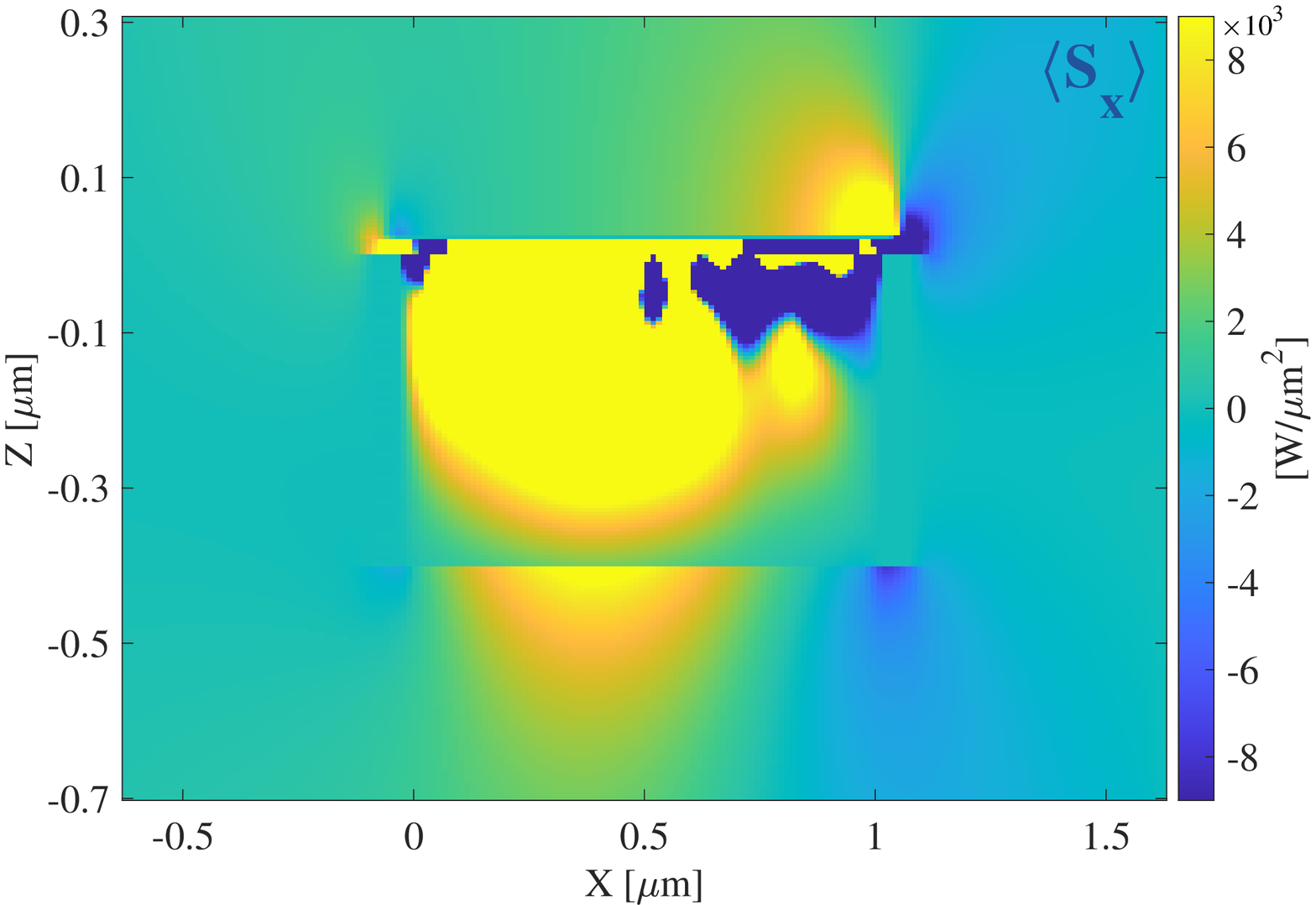}
        \caption{} 
        \label{fig:sx}
    \end{subfigure}
    %\hfill
    \begin{subfigure}{.48\textwidth}
        \includegraphics[width=\linewidth]{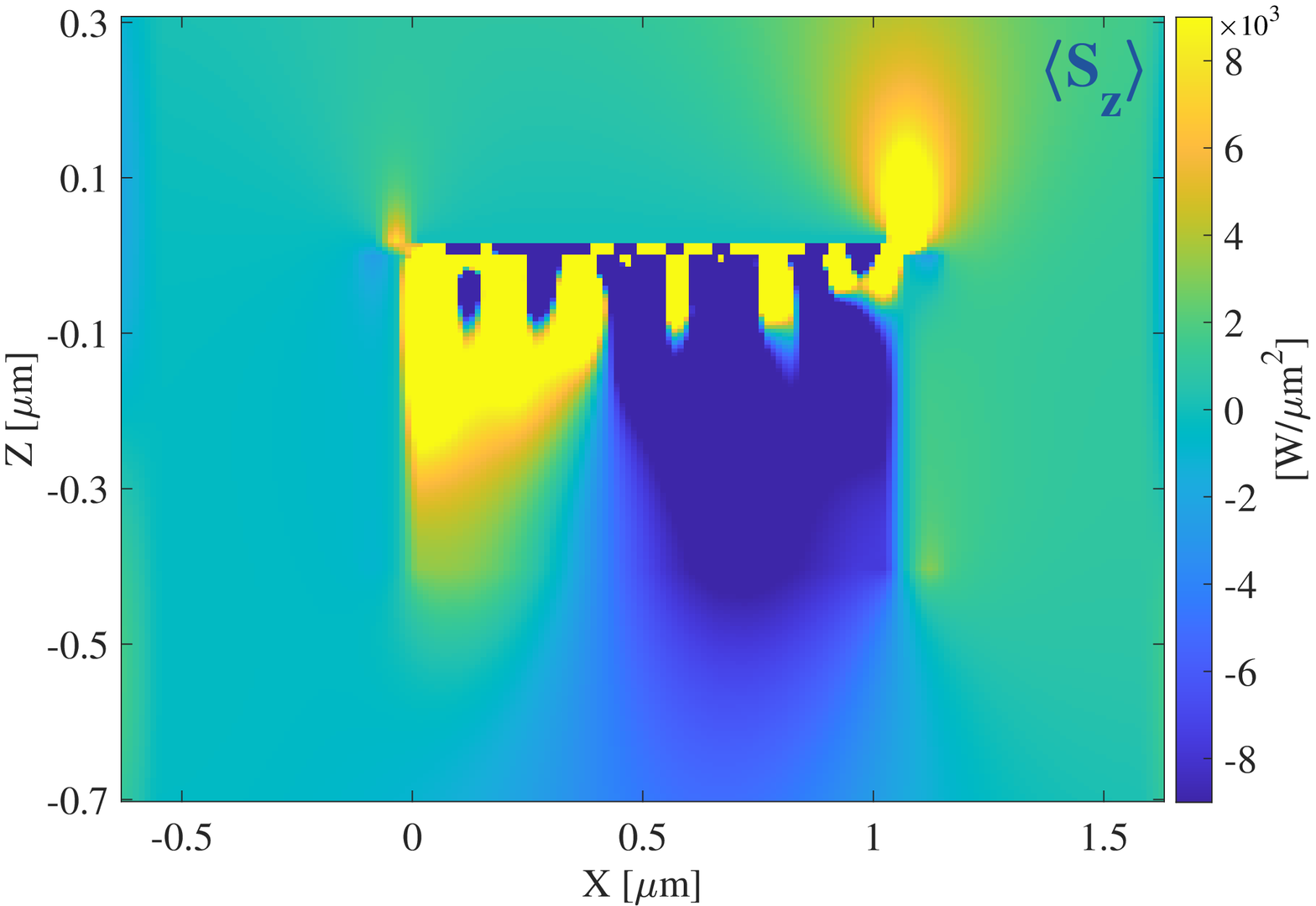}
        \caption{} 
        \label{fig:sz}
    \end{subfigure}
    \caption{Spatial distribution of the components of the time-averaged Poynting vector $\langle S_x \rangle$ (a) and $\langle S_z \rangle$ (b) of the THz EM radiation generated by the graphene transistor structure in the DS instability regime at $n_0 = \SI[per-mode=reciprocal]{4.2e16}{\per\meter\squared}$, $v_0 = \SI{4e5}{\meter\per\second}$, and $\tau = \SI{5}{\pico\second}$.} 
    \label{fig:s-field}
\end{figure}

\begin{figure}[ht]
    \centering
    \includegraphics[width=\linewidth]{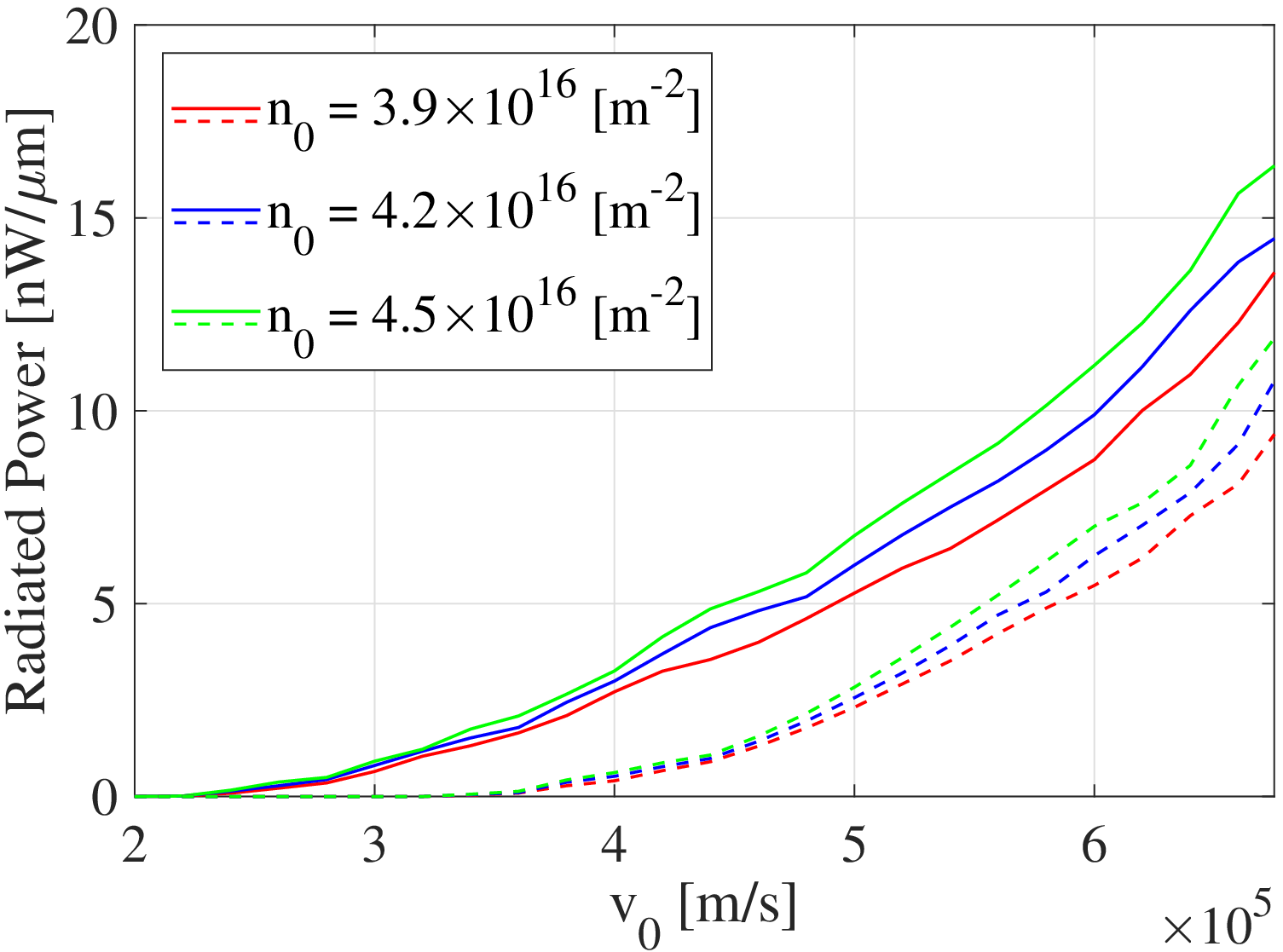}
    \caption{The total radiated power at the instability endpoint as a function of the electron drift velocity $v_0$ at several values of the equilibrium electron density $n_0$. Solid (dashed) lines correspond to the momentum relaxation time $\tau = \SI{5}{\pico\second}$ ($\tau = \SI{2.5}{\pico\second}$).}
    \label{fig:n0}
\end{figure}

The solution of Maxwell's equations, found using the electrodynamic FTDT solver in our Multiphysics Simulation Platform, provides the values of the electric $\bm{E}$ and magnetic $\bm{H}$ components of the EM field radiated by the graphene transistor channel at the instability endpoint. These components were used to find the spatial distribution pattern of the time-averaged components of the Poynting vector $\bm{S}$ in the area surrounding the graphene channel, 
\begin{equation}
    \langle \bm{S} \rangle = \frac{1}{T}\int_{t}^{t+T} \bm{E}(t) \times \bm{H}(t) \,dt. \
\end{equation}
The patterns for the $x$ and $z$ components of the Poynting vector are shown in Fig.~\ref{fig:s-field}(\subref{fig:sx}) and \ref{fig:s-field}(\subref{fig:sz}), respectively. It follows from these patterns that while the metal gate effectively blocks the EM radiation generated mostly in the gap between the gate and the graphene layer, the infinitely thin graphene layer is practically transparent. Also, asymmetric boundaries lead to an uneven EM field distribution near the source and drain contacts. These features determine the resulting radiation pattern. The total EM power $P$ emitted by the graphene transistor can be calculated by integrating the normal component of the Poynting vector over a continuous boundary encircling the device. In Fig.~\ref{fig:n0}, we plotted the total power $P$ as a function of the drift velocity $v_0$ at several different values of the equilibrium electron density $n_0$ and electron momentum relaxation time $\tau$. The available range of $v_0$ is limited by the electron saturation velocity in graphene: $v_0 \lesssim 0.5 v_F$ \cite{meric08}. As expected, the power $P$ increases with the drift velocity because of the increased instability increment, see Fig.~\ref{fig:plasmon}(\subref{fig:gamma}). The larger instability increment results in a larger amplitude of the electron oscillations in the final stationary state. The radiated power strongly depends on the relaxation time $\tau$, increasing at smaller scattering rates. It happens due to the shift of the instability threshold to smaller $v_0$, and also because of the decreased Joule heating in the final stationary state leading to redistribution of the power supplied by the external circuit towards the radiation. The power $P$ only slightly increases with the equilibrium electron density $n_0$ due to increased oscillator strength in the final stationary state. In our one-dimensional model, the calculations yield $P = \SI{4.6}{\nano\watt\per\micro\meter}$ for the values of parameters used in Fig.~\ref{fig:s-field}. If the finite width $W$ of the transistor structure is taken into account, the total power can roughly be estimated as $P = \SI{460}{\nano\watt}$ at $W = \SI{100}{\micro\meter}$. This number can only be considered as a lower bound of the emitted power because our one-dimensional channel model and 2D Maxwell's equations do not give full description of the three-dimensional radiation pattern. These power estimates are comparable with similar estimates for the III-V semiconductor-based transistors \cite{nafari18}. The power radiated at frequency $\omega$ can be evaluated as 
\begin{equation} \label{P}
    P = \frac{1}{2}\oint_\mathcal{C} \operatorname{Re}\left\{\bm{E}(\omega)\times\bm{H}(\omega)\right\}\cdot\mathbf{n}\,dl,
\end{equation}
\noindent
where $\bm{E}(\omega)$ and $\bm{H}(\omega)$ are the Fourier transforms of the electric and magnetic field vectors and $\mathcal{C}$ is an enclosed boundary with normal vector $\mathbf{n}$. For the device parameters used in Fig.~\ref{fig:s-field}, a fundamental frequency of $f_0 = \SI{1.2}{\tera\hertz}$ was calculated. The EM power radiated at this frequency is $P = \SI{3.1}{\nano\watt\per\micro\meter}$, which means that only 67\% of the total power is radiated at the fundamental frequency.

\section{Discussion and Concluding Remarks} \label{con}
The growth of plasma waves due to the DS instability is opposed by various dissipative processes damping the plasma waves such as random electron scattering on phonons and impurities in the transistor channel \cite{DS93}, plasmon damping in the metal contacts at the channel boundaries \cite{kapralov20}, and finite viscosity of the electron fluid in graphene \cite{DS93}. In the state-of-the-art graphene structures the electron mean free path $v_\text{F}\tau$ is of the order of \SI{10}{\micro\meter} at low temperatures. In plasmonic cavities of length $L \approx \SI{1}{\micro\meter}$, the instability increment $\omega'' = \frac{v_0}{2L}$ will exceed the plasma wave decrement due to random electron scattering $\frac{1}{2\tau}$ at the drift velocity $v_0 \gtrsim 0.1v_\text{F}$. These values of the drift velocity are accessible in the experiment \cite{meric08}. As shown in \cite{kapralov20}, the plasma wave decrement due to plasmon damping on the metal leads has the same order of magnitude as the decrement due to random electron scattering and should not pose additional problems for experimental realization of the DS instability. 

In view of the above, the plasma wave damping due to finite viscosity of the electron fluid in graphene emerges as the only significant dissipative process suppressing the DS instability. In the hydrodynamic regime, finite viscosity of the electron fluid can be included into the hydrodynamic equations by replacing the Euler equation with the Navier-Stokes equation, accounting for the dissipative processes in the fluid due to internal friction \cite{rudin11,briskot15,Dmitriev97,cheremisin02,li17,mendl21}. Rigorous numerical solution of the non-linear hydrodynamic equations for viscous electron fluid together with Maxwell's equations is beyond the scope of this paper and will be considered elsewhere \cite{crabb21}. However, some qualitative estimates of the effect of the finite viscosity on the DS instability can be made based on the linear response theory. In the hydrodynamic regime, the plasma wave decrement $\gamma_v$ due to finite viscosity is determined as $\gamma_v \sim \nu k^2$, where $\nu \sim v_\text{F}^2 \tau_\text{ee}$ is the electron fluid viscosity, $k \sim \frac{n}{L}$ is the wave vector of the $n$-th plasma mode in the cavity of length $L$, and $\tau_\text{ee}$ is the time of electron-electron scattering \cite{rudin11,DS93}. The plasma damping due to viscosity increases for the higher order modes with larger wave vector $k$. It also increases with electron-electron scattering time $\tau_\text{ee}$. The upper bound of the viscosity contribution to the damping of the plasma mode of frequency $\omega_n$ can be determined from the condition $\omega_n \tau_\text{ee} \sim 1$. This condition is justified because at larger values of $\tau_\text{ee}$, the plasmonic system experiences a transition from the hydrodynamic regime to the ballistic regime where viscosity correction to the plasmon damping disappears \cite{dmitriev01,mendl18}. For the fundamental plasma mode of frequency $\omega_1 \sim \frac{v_\text{p}}{L}$, the upper bound of the damping due to viscosity is $\gamma_v \sim \frac{v_\text{F}^2}{v_\text{p}L}$, where plasmon velocity $v_\text{p}$ is defined in Eq.~(\ref{vp}). The fundamental mode remains unstable if $\omega'' > \gamma_v$. This condition determines the threshold for the drift velocity $v_0 \gtrsim \frac{v_\text{F}^2}{v_\text{p}}$. For a gated graphene structure with  $n_0 = \SI[per-mode=reciprocal]{1e16}{\per\meter\squared}$, $d = \SI{100}{\nano\meter}$, $\epsilon = 3.8$, we obtain $v_0 \gtrsim 0.15v_\text{F}$. This threshold decreases further with increasing electron density $n_0$. 

The above estimates show that the dissipative plasmon losses in the graphene transistor channel do not present a significant obstacle for observing the DS instability. Yet another non-dissipative process may strongly impact the experimental realization of the DS instability. The instability increment $\omega''$ critically depends on the asymmetry of the boundaries at the opposite ends of the plasmonic cavity formed in the transistor channel \cite{DS93,nafari18,cheremisin99,svintsov18}. In this paper, the instability increment $\omega''$ was derived under the assumption of the ideal asymmetric boundary conditions in Eq.~(\ref{eq25}), which correspond to the zero impedance between the source and gate, $Z_\text{gs} = 0$, and infinite impedance between the drain and gate, $Z_\text{gd} = \infty$. These boundary conditions were first introduced in the pioneering paper by Dyakonov and Shur \cite{DS93}. However, in any real experimental system these capacitive impedances always have some finite values determined by the system geometry and material parameters. The effect of finite $Z_\text{gs}$ and $Z_\text{gd}$ on the DS instability in the semiconductor structures was considered in several papers \cite{nafari18,cheremisin99,svintsov18}. It was shown that any deviations of $Z_\text{gs}$ and $Z_\text{gd}$ from the ideal values change the resonant plasma frequencies in the cavity \cite{nafari18}, and most importantly, decrease the instability increment \cite{nafari18,cheremisin99}. In the limit $Z_\text{gs} = Z_\text{gd}$, the instability increment turns to zero. The lack of sufficient asymmetry of the boundaries may suppress the DS instability, so special care should be taken to satisfy condition $Z_\text{gs} \ll Z_\text{gd}$ in the experimental studies.

In summary, we presented comprehensive, analytical, and numerical studies of the DS instability in graphene transistors in the hydrodynamic regime. We analyzed conditions necessary for the onset of the DS instability, the properties of the final stationary state (the instability endpoint) of the electron fluid in the graphene transistor channel, as well as THz EM radiation emitted in this state. The developed Multiphysics Simulation Platform allowed us to numerically solve the non-linear hydrodynamic equations together with the electrodynamics equations providing the powerful and versatile tool for future studies of similar systems. We demonstrated the feasibility of the DS instability in the current state-of-the-art graphene transistors which can be used for designing an on-chip tunable THz plasmonic generator with potential applications ranging from the short distance communications in the wireless on-chip networks to the novel imaging and sensing techniques.

\section{Acknowledgements}
This work was partially supported by the Air Force Office of Scientific Research (AFOSR) under Grant no. FA9550-16-1-0188, and the US National Science Foundation under Grant no. CNS-2011411.
\vfill

\bibliography{mybib}

%apsrev4-2.bst 2019-01-14 (MD) hand-edited version of apsrev4-1.bst
%Control: key (0)
%Control: author (8) initials jnrlst
%Control: editor formatted (1) identically to author
%Control: production of article title (0) allowed
%Control: page (0) single
%Control: year (1) truncated
%Control: production of eprint (0) enabled
\begin{thebibliography}{56}%
\makeatletter
\providecommand \@ifxundefined [1]{%
 \@ifx{#1\undefined}
}%
\providecommand \@ifnum [1]{%
 \ifnum #1\expandafter \@firstoftwo
 \else \expandafter \@secondoftwo
 \fi
}%
\providecommand \@ifx [1]{%
 \ifx #1\expandafter \@firstoftwo
 \else \expandafter \@secondoftwo
 \fi
}%
\providecommand \natexlab [1]{#1}%
\providecommand \enquote  [1]{``#1''}%
\providecommand \bibnamefont  [1]{#1}%
\providecommand \bibfnamefont [1]{#1}%
\providecommand \citenamefont [1]{#1}%
\providecommand \href@noop [0]{\@secondoftwo}%
\providecommand \href [0]{\begingroup \@sanitize@url \@href}%
\providecommand \@href[1]{\@@startlink{#1}\@@href}%
\providecommand \@@href[1]{\endgroup#1\@@endlink}%
\providecommand \@sanitize@url [0]{\catcode `\\12\catcode `\$12\catcode
  `\&12\catcode `\#12\catcode `\^12\catcode `\_12\catcode `\%12\relax}%
\providecommand \@@startlink[1]{}%
\providecommand \@@endlink[0]{}%
\providecommand \url  [0]{\begingroup\@sanitize@url \@url }%
\providecommand \@url [1]{\endgroup\@href {#1}{\urlprefix }}%
\providecommand \urlprefix  [0]{URL }%
\providecommand \Eprint [0]{\href }%
\providecommand \doibase [0]{https://doi.org/}%
\providecommand \selectlanguage [0]{\@gobble}%
\providecommand \bibinfo  [0]{\@secondoftwo}%
\providecommand \bibfield  [0]{\@secondoftwo}%
\providecommand \translation [1]{[#1]}%
\providecommand \BibitemOpen [0]{}%
\providecommand \bibitemStop [0]{}%
\providecommand \bibitemNoStop [0]{.\EOS\space}%
\providecommand \EOS [0]{\spacefactor3000\relax}%
\providecommand \BibitemShut  [1]{\csname bibitem#1\endcsname}%
\let\auto@bib@innerbib\@empty
%</preamble>
\bibitem [{\citenamefont {Fitch}\ and\ \citenamefont
  {Osiander}(2004)}]{Sensing04}%
  \BibitemOpen
  \bibfield  {author} {\bibinfo {author} {\bibfnamefont {M.}~\bibnamefont
  {Fitch}}\ and\ \bibinfo {author} {\bibfnamefont {R.}~\bibnamefont
  {Osiander}},\ }\bibfield  {title} {\bibinfo {title} {Terahertz waves for
  communications and sensing},\ }\href@noop {} {\bibfield  {journal} {\bibinfo
  {journal} {Johns Hopkins APL Tech Dig}\ }\textbf {\bibinfo {volume} {25}}
  (\bibinfo {year} {2004})}\BibitemShut {NoStop}%
\bibitem [{\citenamefont {Federici}\ \emph {et~al.}(2005)\citenamefont
  {Federici}, \citenamefont {Schulkin}, \citenamefont {Huang}, \citenamefont
  {Gary}, \citenamefont {Barat}, \citenamefont {Oliveira},\ and\ \citenamefont
  {Zimdars}}]{federici05}%
  \BibitemOpen
  \bibfield  {author} {\bibinfo {author} {\bibfnamefont {J.~F.}\ \bibnamefont
  {Federici}}, \bibinfo {author} {\bibfnamefont {B.}~\bibnamefont {Schulkin}},
  \bibinfo {author} {\bibfnamefont {F.}~\bibnamefont {Huang}}, \bibinfo
  {author} {\bibfnamefont {D.}~\bibnamefont {Gary}}, \bibinfo {author}
  {\bibfnamefont {R.}~\bibnamefont {Barat}}, \bibinfo {author} {\bibfnamefont
  {F.}~\bibnamefont {Oliveira}},\ and\ \bibinfo {author} {\bibfnamefont
  {D.}~\bibnamefont {Zimdars}},\ }\bibfield  {title} {\bibinfo {title} {Thz
  imaging and sensing for security applications—explosives, weapons and
  drugs},\ }\href@noop {} {\bibfield  {journal} {\bibinfo  {journal} {Semicon.
  Sci. and Technol.}\ }\textbf {\bibinfo {volume} {20}},\ \bibinfo {pages}
  {S266} (\bibinfo {year} {2005})}\BibitemShut {NoStop}%
\bibitem [{\citenamefont {{Jornet}}\ and\ \citenamefont
  {{Akyildiz}}(2011)}]{Channel11}%
  \BibitemOpen
  \bibfield  {author} {\bibinfo {author} {\bibfnamefont {J.~M.}\ \bibnamefont
  {{Jornet}}}\ and\ \bibinfo {author} {\bibfnamefont {I.~F.}\ \bibnamefont
  {{Akyildiz}}},\ }\bibfield  {title} {\bibinfo {title} {Channel modeling and
  capacity analysis for electromagnetic wireless nanonetworks in the terahertz
  band},\ }\href {https://doi.org/10.1109/TWC.2011.081011.100545} {\bibfield
  {journal} {\bibinfo  {journal} {IEEE Trans. Wirel. Commun.}\ }\textbf
  {\bibinfo {volume} {10}},\ \bibinfo {pages} {3211} (\bibinfo {year}
  {2011})}\BibitemShut {NoStop}%
\bibitem [{\citenamefont {Akyildiz}\ \emph
  {et~al.}(2014{\natexlab{a}})\citenamefont {Akyildiz}, \citenamefont
  {Jornet},\ and\ \citenamefont {Han}}]{akyildiz14}%
  \BibitemOpen
  \bibfield  {author} {\bibinfo {author} {\bibfnamefont {I.~F.}\ \bibnamefont
  {Akyildiz}}, \bibinfo {author} {\bibfnamefont {J.~M.}\ \bibnamefont
  {Jornet}},\ and\ \bibinfo {author} {\bibfnamefont {C.}~\bibnamefont {Han}},\
  }\bibfield  {title} {\bibinfo {title} {Terahertz band: Next frontier for
  wireless communications},\ }\href@noop {} {\bibfield  {journal} {\bibinfo
  {journal} {Phys Commun.}\ }\textbf {\bibinfo {volume} {12}},\ \bibinfo
  {pages} {16} (\bibinfo {year} {2014}{\natexlab{a}})}\BibitemShut {NoStop}%
\bibitem [{\citenamefont {Mittleman}(2017)}]{mittleman17}%
  \BibitemOpen
  \bibfield  {author} {\bibinfo {author} {\bibfnamefont {D.~M.}\ \bibnamefont
  {Mittleman}},\ }\bibfield  {title} {\bibinfo {title} {Perspective: Terahertz
  science and technology},\ }\href@noop {} {\bibfield  {journal} {\bibinfo
  {journal} {J. Appl. Phys.}\ }\textbf {\bibinfo {volume} {122}},\ \bibinfo
  {pages} {230901} (\bibinfo {year} {2017})}\BibitemShut {NoStop}%
\bibitem [{\citenamefont {Akyildiz}\ \emph
  {et~al.}(2014{\natexlab{b}})\citenamefont {Akyildiz}, \citenamefont
  {Jornet},\ and\ \citenamefont {Han}}]{akyildiz14teranets}%
  \BibitemOpen
  \bibfield  {author} {\bibinfo {author} {\bibfnamefont {I.~F.}\ \bibnamefont
  {Akyildiz}}, \bibinfo {author} {\bibfnamefont {J.~M.}\ \bibnamefont
  {Jornet}},\ and\ \bibinfo {author} {\bibfnamefont {C.}~\bibnamefont {Han}},\
  }\bibfield  {title} {\bibinfo {title} {Teranets: Ultra-broadband
  communication networks in the terahertz band},\ }\href@noop {} {\bibfield
  {journal} {\bibinfo  {journal} {IEEE Wirel. Commun.}\ }\textbf {\bibinfo
  {volume} {21}},\ \bibinfo {pages} {130} (\bibinfo {year}
  {2014}{\natexlab{b}})}\BibitemShut {NoStop}%
\bibitem [{\citenamefont {Polese}\ \emph {et~al.}(2020)\citenamefont {Polese},
  \citenamefont {Jornet}, \citenamefont {Melodia},\ and\ \citenamefont
  {Zorzi}}]{polese20}%
  \BibitemOpen
  \bibfield  {author} {\bibinfo {author} {\bibfnamefont {M.}~\bibnamefont
  {Polese}}, \bibinfo {author} {\bibfnamefont {J.~M.}\ \bibnamefont {Jornet}},
  \bibinfo {author} {\bibfnamefont {T.}~\bibnamefont {Melodia}},\ and\ \bibinfo
  {author} {\bibfnamefont {M.}~\bibnamefont {Zorzi}},\ }\bibfield  {title}
  {\bibinfo {title} {Toward end-to-end, full-stack 6g terahertz networks},\
  }\href@noop {} {\bibfield  {journal} {\bibinfo  {journal} {IEEE Commun.
  Mag.}\ }\textbf {\bibinfo {volume} {58}},\ \bibinfo {pages} {48} (\bibinfo
  {year} {2020})}\BibitemShut {NoStop}%
\bibitem [{\citenamefont {Israel}\ \emph {et~al.}(2013)\citenamefont {Israel},
  \citenamefont {Martinovic}, \citenamefont {Fischer}, \citenamefont
  {Jenning},\ and\ \citenamefont {Landau}}]{israel13}%
  \BibitemOpen
  \bibfield  {author} {\bibinfo {author} {\bibfnamefont {J.}~\bibnamefont
  {Israel}}, \bibinfo {author} {\bibfnamefont {J.}~\bibnamefont {Martinovic}},
  \bibinfo {author} {\bibfnamefont {A.}~\bibnamefont {Fischer}}, \bibinfo
  {author} {\bibfnamefont {M.}~\bibnamefont {Jenning}},\ and\ \bibinfo {author}
  {\bibfnamefont {L.}~\bibnamefont {Landau}},\ }\bibfield  {title} {\bibinfo
  {title} {Optimal antenna positioning for wireless board-to-board
  communication using a butler matrix beamforming network},\ }in\ \href@noop {}
  {\emph {\bibinfo {booktitle} {WSA 2013; 17th International ITG Workshop on
  Smart Antennas}}}\ (\bibinfo {organization} {VDE},\ \bibinfo {year} {2013})\
  pp.\ \bibinfo {pages} {1--7}\BibitemShut {NoStop}%
\bibitem [{\citenamefont {Moltchanov}\ \emph {et~al.}(2016)\citenamefont
  {Moltchanov}, \citenamefont {Antonov}, \citenamefont {Kluchev}, \citenamefont
  {Borunova}, \citenamefont {Kustarev}, \citenamefont {Petrov}, \citenamefont
  {Koucheryavy},\ and\ \citenamefont {Platunov}}]{moltchanov16}%
  \BibitemOpen
  \bibfield  {author} {\bibinfo {author} {\bibfnamefont {D.}~\bibnamefont
  {Moltchanov}}, \bibinfo {author} {\bibfnamefont {A.}~\bibnamefont {Antonov}},
  \bibinfo {author} {\bibfnamefont {A.}~\bibnamefont {Kluchev}}, \bibinfo
  {author} {\bibfnamefont {K.}~\bibnamefont {Borunova}}, \bibinfo {author}
  {\bibfnamefont {P.}~\bibnamefont {Kustarev}}, \bibinfo {author}
  {\bibfnamefont {V.}~\bibnamefont {Petrov}}, \bibinfo {author} {\bibfnamefont
  {Y.}~\bibnamefont {Koucheryavy}},\ and\ \bibinfo {author} {\bibfnamefont
  {A.}~\bibnamefont {Platunov}},\ }\bibfield  {title} {\bibinfo {title}
  {Statistical traffic properties and model inference for shared cache
  interface in multi-core cpus},\ }\href@noop {} {\bibfield  {journal}
  {\bibinfo  {journal} {IEEE Access}\ }\textbf {\bibinfo {volume} {4}},\
  \bibinfo {pages} {4829} (\bibinfo {year} {2016})}\BibitemShut {NoStop}%
\bibitem [{\citenamefont {Akyildiz}\ and\ \citenamefont
  {Jornet}(2016)}]{akyildiz16}%
  \BibitemOpen
  \bibfield  {author} {\bibinfo {author} {\bibfnamefont {I.~F.}\ \bibnamefont
  {Akyildiz}}\ and\ \bibinfo {author} {\bibfnamefont {J.~M.}\ \bibnamefont
  {Jornet}},\ }\bibfield  {title} {\bibinfo {title} {Realizing ultra-massive
  mimo (1024$\times$ 1024) communication in the (0.06--10) terahertz band},\
  }\href@noop {} {\bibfield  {journal} {\bibinfo  {journal} {Nano Commun.
  Netw.}\ }\textbf {\bibinfo {volume} {8}},\ \bibinfo {pages} {46} (\bibinfo
  {year} {2016})}\BibitemShut {NoStop}%
\bibitem [{\citenamefont {Dyakonov}\ and\ \citenamefont {Shur}(1993)}]{DS93}%
  \BibitemOpen
  \bibfield  {author} {\bibinfo {author} {\bibfnamefont {M.}~\bibnamefont
  {Dyakonov}}\ and\ \bibinfo {author} {\bibfnamefont {M.~S.}\ \bibnamefont
  {Shur}},\ }\bibfield  {title} {\bibinfo {title} {Shallow water analogy for a
  ballistic field effect transistor: New mechanism of plasma wave generation by
  dc current},\ }\href {https://doi.org/10.1103/PhysRevLett.71.2465} {\bibfield
   {journal} {\bibinfo  {journal} {Phys. Rev. Lett.}\ }\textbf {\bibinfo
  {volume} {71}},\ \bibinfo {pages} {2465} (\bibinfo {year}
  {1993})}\BibitemShut {NoStop}%
\bibitem [{\citenamefont {Ryzhii}\ \emph {et~al.}(2005)\citenamefont {Ryzhii},
  \citenamefont {Satou},\ and\ \citenamefont {Shur}}]{ryzhii05}%
  \BibitemOpen
  \bibfield  {author} {\bibinfo {author} {\bibfnamefont {V.}~\bibnamefont
  {Ryzhii}}, \bibinfo {author} {\bibfnamefont {A.}~\bibnamefont {Satou}},\ and\
  \bibinfo {author} {\bibfnamefont {M.~S.}\ \bibnamefont {Shur}},\ }\bibfield
  {title} {\bibinfo {title} {Transit-time mechanism of plasma instability in
  high electron mobility transistors},\ }\href@noop {} {\bibfield  {journal}
  {\bibinfo  {journal} {Phys. Status Solidi A}\ }\textbf {\bibinfo {volume}
  {202}},\ \bibinfo {pages} {R113} (\bibinfo {year} {2005})}\BibitemShut
  {NoStop}%
\bibitem [{\citenamefont {Mikhailov}(1998)}]{mikhailov98}%
  \BibitemOpen
  \bibfield  {author} {\bibinfo {author} {\bibfnamefont {S.~A.}\ \bibnamefont
  {Mikhailov}},\ }\bibfield  {title} {\bibinfo {title} {Plasma instability and
  amplification of electromagnetic waves in low-dimensional electron systems},\
  }\href@noop {} {\bibfield  {journal} {\bibinfo  {journal} {Phys. Rev. B}\
  }\textbf {\bibinfo {volume} {58}},\ \bibinfo {pages} {1517} (\bibinfo {year}
  {1998})}\BibitemShut {NoStop}%
\bibitem [{\citenamefont {Petrov}\ \emph {et~al.}(2017)\citenamefont {Petrov},
  \citenamefont {Svintsov}, \citenamefont {Ryzhii},\ and\ \citenamefont
  {Shur}}]{petrov17}%
  \BibitemOpen
  \bibfield  {author} {\bibinfo {author} {\bibfnamefont {A.~S.}\ \bibnamefont
  {Petrov}}, \bibinfo {author} {\bibfnamefont {D.}~\bibnamefont {Svintsov}},
  \bibinfo {author} {\bibfnamefont {V.}~\bibnamefont {Ryzhii}},\ and\ \bibinfo
  {author} {\bibfnamefont {M.~S.}\ \bibnamefont {Shur}},\ }\bibfield  {title}
  {\bibinfo {title} {Amplified-reflection plasmon instabilities in grating-gate
  plasmonic crystals},\ }\href@noop {} {\bibfield  {journal} {\bibinfo
  {journal} {Phys. Rev. B}\ }\textbf {\bibinfo {volume} {95}},\ \bibinfo
  {pages} {045405} (\bibinfo {year} {2017})}\BibitemShut {NoStop}%
\bibitem [{\citenamefont {Aizin}\ \emph {et~al.}(2020)\citenamefont {Aizin},
  \citenamefont {Mikalopas},\ and\ \citenamefont {Shur}}]{aizin20}%
  \BibitemOpen
  \bibfield  {author} {\bibinfo {author} {\bibfnamefont {G.~R.}\ \bibnamefont
  {Aizin}}, \bibinfo {author} {\bibfnamefont {J.}~\bibnamefont {Mikalopas}},\
  and\ \bibinfo {author} {\bibfnamefont {M.~S.}\ \bibnamefont {Shur}},\
  }\bibfield  {title} {\bibinfo {title} {Plasmonic instabilities in
  two-dimensional electron channels of variable width},\ }\href@noop {}
  {\bibfield  {journal} {\bibinfo  {journal} {Phys. Rev. B}\ }\textbf {\bibinfo
  {volume} {101}},\ \bibinfo {pages} {245404} (\bibinfo {year}
  {2020})}\BibitemShut {NoStop}%
\bibitem [{\citenamefont {Kachorovskii}\ and\ \citenamefont
  {Shur}(2012)}]{kachorovskii12}%
  \BibitemOpen
  \bibfield  {author} {\bibinfo {author} {\bibfnamefont {V.~Y.}\ \bibnamefont
  {Kachorovskii}}\ and\ \bibinfo {author} {\bibfnamefont {M.~S.}\ \bibnamefont
  {Shur}},\ }\bibfield  {title} {\bibinfo {title} {Current-induced terahertz
  oscillations in plasmonic crystal},\ }\href@noop {} {\bibfield  {journal}
  {\bibinfo  {journal} {Appl. Phys. Lett.}\ }\textbf {\bibinfo {volume}
  {100}},\ \bibinfo {pages} {232108} (\bibinfo {year} {2012})}\BibitemShut
  {NoStop}%
\bibitem [{\citenamefont {Aizin}\ \emph {et~al.}(2016)\citenamefont {Aizin},
  \citenamefont {Mikalopas},\ and\ \citenamefont {Shur}}]{aizin16}%
  \BibitemOpen
  \bibfield  {author} {\bibinfo {author} {\bibfnamefont {G.~R.}\ \bibnamefont
  {Aizin}}, \bibinfo {author} {\bibfnamefont {J.}~\bibnamefont {Mikalopas}},\
  and\ \bibinfo {author} {\bibfnamefont {M.~S.}\ \bibnamefont {Shur}},\
  }\bibfield  {title} {\bibinfo {title} {Current-driven plasmonic boom
  instability in three-dimensional gated periodic ballistic nanostructures},\
  }\href@noop {} {\bibfield  {journal} {\bibinfo  {journal} {Phys. Rev. B}\
  }\textbf {\bibinfo {volume} {93}},\ \bibinfo {pages} {195315} (\bibinfo
  {year} {2016})}\BibitemShut {NoStop}%
\bibitem [{\citenamefont {Knap}\ \emph {et~al.}(2004)\citenamefont {Knap},
  \citenamefont {{\L}usakowski}, \citenamefont {Parenty}, \citenamefont
  {Bollaert}, \citenamefont {Cappy}, \citenamefont {Popov},\ and\ \citenamefont
  {Shur}}]{knap04}%
  \BibitemOpen
  \bibfield  {author} {\bibinfo {author} {\bibfnamefont {W.}~\bibnamefont
  {Knap}}, \bibinfo {author} {\bibfnamefont {J.}~\bibnamefont {{\L}usakowski}},
  \bibinfo {author} {\bibfnamefont {T.}~\bibnamefont {Parenty}}, \bibinfo
  {author} {\bibfnamefont {S.}~\bibnamefont {Bollaert}}, \bibinfo {author}
  {\bibfnamefont {A.}~\bibnamefont {Cappy}}, \bibinfo {author} {\bibfnamefont
  {V.~V.}\ \bibnamefont {Popov}},\ and\ \bibinfo {author} {\bibfnamefont
  {M.~S.}\ \bibnamefont {Shur}},\ }\bibfield  {title} {\bibinfo {title}
  {Terahertz emission by plasma waves in 60 nm gate high electron mobility
  transistors},\ }\href@noop {} {\bibfield  {journal} {\bibinfo  {journal}
  {Appl. Phys. Lett.}\ }\textbf {\bibinfo {volume} {84}},\ \bibinfo {pages}
  {2331} (\bibinfo {year} {2004})}\BibitemShut {NoStop}%
\bibitem [{\citenamefont {{\L}usakowski}\ \emph {et~al.}(2005)\citenamefont
  {{\L}usakowski}, \citenamefont {Knap}, \citenamefont {Dyakonova},
  \citenamefont {Varani}, \citenamefont {Mateos}, \citenamefont {Gonzalez},
  \citenamefont {Roelens}, \citenamefont {Bollaert}, \citenamefont {Cappy},\
  and\ \citenamefont {Karpierz}}]{lusakowski05}%
  \BibitemOpen
  \bibfield  {author} {\bibinfo {author} {\bibfnamefont {J.}~\bibnamefont
  {{\L}usakowski}}, \bibinfo {author} {\bibfnamefont {W.}~\bibnamefont {Knap}},
  \bibinfo {author} {\bibfnamefont {N.}~\bibnamefont {Dyakonova}}, \bibinfo
  {author} {\bibfnamefont {L.}~\bibnamefont {Varani}}, \bibinfo {author}
  {\bibfnamefont {J.}~\bibnamefont {Mateos}}, \bibinfo {author} {\bibfnamefont
  {T.}~\bibnamefont {Gonzalez}}, \bibinfo {author} {\bibfnamefont
  {Y.}~\bibnamefont {Roelens}}, \bibinfo {author} {\bibfnamefont
  {S.}~\bibnamefont {Bollaert}}, \bibinfo {author} {\bibfnamefont
  {A.}~\bibnamefont {Cappy}},\ and\ \bibinfo {author} {\bibfnamefont
  {K.}~\bibnamefont {Karpierz}},\ }\bibfield  {title} {\bibinfo {title}
  {Voltage tuneable terahertz emission from a ballistic nanometer ingaas/
  inalas transistor},\ }\href@noop {} {\bibfield  {journal} {\bibinfo
  {journal} {J. Appl. Phys.}\ }\textbf {\bibinfo {volume} {97}},\ \bibinfo
  {pages} {064307} (\bibinfo {year} {2005})}\BibitemShut {NoStop}%
\bibitem [{\citenamefont {Dyakonova}\ \emph {et~al.}(2005)\citenamefont
  {Dyakonova}, \citenamefont {Teppe}, \citenamefont {{\L}usakowski},
  \citenamefont {Knap}, \citenamefont {Levinshtein}, \citenamefont {Dmitriev},
  \citenamefont {Shur}, \citenamefont {Bollaert},\ and\ \citenamefont
  {Cappy}}]{dyakonova05}%
  \BibitemOpen
  \bibfield  {author} {\bibinfo {author} {\bibfnamefont {N.}~\bibnamefont
  {Dyakonova}}, \bibinfo {author} {\bibfnamefont {F.}~\bibnamefont {Teppe}},
  \bibinfo {author} {\bibfnamefont {J.}~\bibnamefont {{\L}usakowski}}, \bibinfo
  {author} {\bibfnamefont {W.}~\bibnamefont {Knap}}, \bibinfo {author}
  {\bibfnamefont {M.}~\bibnamefont {Levinshtein}}, \bibinfo {author}
  {\bibfnamefont {A.~P.}\ \bibnamefont {Dmitriev}}, \bibinfo {author}
  {\bibfnamefont {M.~S.}\ \bibnamefont {Shur}}, \bibinfo {author}
  {\bibfnamefont {S.}~\bibnamefont {Bollaert}},\ and\ \bibinfo {author}
  {\bibfnamefont {A.}~\bibnamefont {Cappy}},\ }\bibfield  {title} {\bibinfo
  {title} {Magnetic field effect on the terahertz emission from nanometer
  ingaas/alinas high electron mobility transistors},\ }\href@noop {} {\bibfield
   {journal} {\bibinfo  {journal} {J. Appl. Phys.}\ }\textbf {\bibinfo {volume}
  {97}},\ \bibinfo {pages} {114313} (\bibinfo {year} {2005})}\BibitemShut
  {NoStop}%
\bibitem [{\citenamefont {Dyakonova}\ \emph {et~al.}(2006)\citenamefont
  {Dyakonova}, \citenamefont {Fatimy}, \citenamefont {{\L}usakowski},
  \citenamefont {Dyakonov}, \citenamefont {Poisson}, \citenamefont {Morvan},
  \citenamefont {Bollaert}, \citenamefont {Shchepetov}, \citenamefont {Roelens}
  \emph {et~al.}}]{dyakonova06}%
  \BibitemOpen
  \bibfield  {author} {\bibinfo {author} {\bibfnamefont {N.}~\bibnamefont
  {Dyakonova}}, \bibinfo {author} {\bibfnamefont {A.~E.}\ \bibnamefont
  {Fatimy}}, \bibinfo {author} {\bibfnamefont {J.}~\bibnamefont
  {{\L}usakowski}}, \bibinfo {author} {\bibfnamefont {W.~K.~M.}\ \bibnamefont
  {Dyakonov}}, \bibinfo {author} {\bibfnamefont {M.~A.}\ \bibnamefont
  {Poisson}}, \bibinfo {author} {\bibfnamefont {E.}~\bibnamefont {Morvan}},
  \bibinfo {author} {\bibfnamefont {S.}~\bibnamefont {Bollaert}}, \bibinfo
  {author} {\bibfnamefont {A.}~\bibnamefont {Shchepetov}}, \bibinfo {author}
  {\bibfnamefont {Y.}~\bibnamefont {Roelens}}, \emph {et~al.},\ }\bibfield
  {title} {\bibinfo {title} {Room-temperature terahertz emission from nanometer
  field-effect transistors},\ }\href@noop {} {\bibfield  {journal} {\bibinfo
  {journal} {Appl. Phys. Lett.}\ }\textbf {\bibinfo {volume} {88}},\ \bibinfo
  {pages} {141906} (\bibinfo {year} {2006})}\BibitemShut {NoStop}%
\bibitem [{\citenamefont {Boubanga-Tombet}\ \emph {et~al.}(2010)\citenamefont
  {Boubanga-Tombet}, \citenamefont {Teppe}, \citenamefont {Torres},
  \citenamefont {Moutaouakil}, \citenamefont {Coquillat}, \citenamefont
  {Dyakonova}, \citenamefont {Consejo}, \citenamefont {Arcade}, \citenamefont
  {Nouvel}, \citenamefont {Marinchio} \emph {et~al.}}]{boubanga10}%
  \BibitemOpen
  \bibfield  {author} {\bibinfo {author} {\bibfnamefont {S.~A.}\ \bibnamefont
  {Boubanga-Tombet}}, \bibinfo {author} {\bibfnamefont {F.}~\bibnamefont
  {Teppe}}, \bibinfo {author} {\bibfnamefont {J.}~\bibnamefont {Torres}},
  \bibinfo {author} {\bibfnamefont {A.~E.}\ \bibnamefont {Moutaouakil}},
  \bibinfo {author} {\bibfnamefont {D.}~\bibnamefont {Coquillat}}, \bibinfo
  {author} {\bibfnamefont {N.}~\bibnamefont {Dyakonova}}, \bibinfo {author}
  {\bibfnamefont {C.}~\bibnamefont {Consejo}}, \bibinfo {author} {\bibfnamefont
  {P.}~\bibnamefont {Arcade}}, \bibinfo {author} {\bibfnamefont
  {P.}~\bibnamefont {Nouvel}}, \bibinfo {author} {\bibfnamefont
  {H.}~\bibnamefont {Marinchio}}, \emph {et~al.},\ }\bibfield  {title}
  {\bibinfo {title} {Room temperature coherent and voltage tunable terahertz
  emission from nanometer-sized field effect transistors},\ }\href@noop {}
  {\bibfield  {journal} {\bibinfo  {journal} {Appl. Phys. Lett.}\ }\textbf
  {\bibinfo {volume} {97}},\ \bibinfo {pages} {262108} (\bibinfo {year}
  {2010})}\BibitemShut {NoStop}%
\bibitem [{\citenamefont {Fatimy}\ \emph {et~al.}(2010)\citenamefont {Fatimy},
  \citenamefont {Dyakonova}, \citenamefont {Meziani}, \citenamefont {Otsuji},
  \citenamefont {Knap}, \citenamefont {Vandenbrouk}, \citenamefont {Madjour},
  \citenamefont {Théron}, \citenamefont {Gaquiere}, \citenamefont {Poisson}
  \emph {et~al.}}]{GaN}%
  \BibitemOpen
  \bibfield  {author} {\bibinfo {author} {\bibfnamefont {A.~E.}\ \bibnamefont
  {Fatimy}}, \bibinfo {author} {\bibfnamefont {N.}~\bibnamefont {Dyakonova}},
  \bibinfo {author} {\bibfnamefont {Y.}~\bibnamefont {Meziani}}, \bibinfo
  {author} {\bibfnamefont {T.}~\bibnamefont {Otsuji}}, \bibinfo {author}
  {\bibfnamefont {W.}~\bibnamefont {Knap}}, \bibinfo {author} {\bibfnamefont
  {S.}~\bibnamefont {Vandenbrouk}}, \bibinfo {author} {\bibfnamefont
  {K.}~\bibnamefont {Madjour}}, \bibinfo {author} {\bibfnamefont
  {D.}~\bibnamefont {Théron}}, \bibinfo {author} {\bibfnamefont
  {C.}~\bibnamefont {Gaquiere}}, \bibinfo {author} {\bibfnamefont {M.~A.}\
  \bibnamefont {Poisson}}, \emph {et~al.},\ }\bibfield  {title} {\bibinfo
  {title} {Algan/gan high electron mobility transistors as a voltage-tunable
  room temperature terahertz sources},\ }\href
  {https://doi.org/10.1063/1.3291101} {\bibfield  {journal} {\bibinfo
  {journal} {J. Appl. Phys.}\ }\textbf {\bibinfo {volume} {107}},\ \bibinfo
  {pages} {024504} (\bibinfo {year} {2010})}\BibitemShut {NoStop}%
\bibitem [{\citenamefont {Onishi}\ \emph {et~al.}(2010)\citenamefont {Onishi},
  \citenamefont {Tanigawa},\ and\ \citenamefont {Takigawa}}]{onishi10}%
  \BibitemOpen
  \bibfield  {author} {\bibinfo {author} {\bibfnamefont {T.}~\bibnamefont
  {Onishi}}, \bibinfo {author} {\bibfnamefont {T.}~\bibnamefont {Tanigawa}},\
  and\ \bibinfo {author} {\bibfnamefont {S.}~\bibnamefont {Takigawa}},\
  }\bibfield  {title} {\bibinfo {title} {High power terahertz emission from a
  single gate algan/gan field effect transistor with periodic ohmic contacts
  for plasmon coupling},\ }\href@noop {} {\bibfield  {journal} {\bibinfo
  {journal} {Appl. Phys. Lett.}\ }\textbf {\bibinfo {volume} {97}},\ \bibinfo
  {pages} {092117} (\bibinfo {year} {2010})}\BibitemShut {NoStop}%
\bibitem [{\citenamefont {Otsuji}\ \emph {et~al.}(2013)\citenamefont {Otsuji},
  \citenamefont {Watanabe}, \citenamefont {Boubanga-Tombet}, \citenamefont
  {Satou}, \citenamefont {Ryzhii}, \citenamefont {Popov},\ and\ \citenamefont
  {Knap}}]{otsuji13}%
  \BibitemOpen
  \bibfield  {author} {\bibinfo {author} {\bibfnamefont {T.}~\bibnamefont
  {Otsuji}}, \bibinfo {author} {\bibfnamefont {T.}~\bibnamefont {Watanabe}},
  \bibinfo {author} {\bibfnamefont {S.~A.}\ \bibnamefont {Boubanga-Tombet}},
  \bibinfo {author} {\bibfnamefont {A.}~\bibnamefont {Satou}}, \bibinfo
  {author} {\bibfnamefont {V.}~\bibnamefont {Ryzhii}}, \bibinfo {author}
  {\bibfnamefont {V.~V.}\ \bibnamefont {Popov}},\ and\ \bibinfo {author}
  {\bibfnamefont {W.}~\bibnamefont {Knap}},\ }\bibfield  {title} {\bibinfo
  {title} {Emission and detection of terahertz radiation using two-dimensional
  plasmons in semiconductor nanoheterostructures for nondestructive
  evaluations},\ }\href@noop {} {\bibfield  {journal} {\bibinfo  {journal}
  {Opt. Eng.}\ }\textbf {\bibinfo {volume} {53}},\ \bibinfo {pages} {031206}
  (\bibinfo {year} {2013})}\BibitemShut {NoStop}%
\bibitem [{\citenamefont {Jak{\v{s}}tas}\ \emph {et~al.}(2017)\citenamefont
  {Jak{\v{s}}tas}, \citenamefont {Grigelionis}, \citenamefont {Janonis},
  \citenamefont {Valu{\v{s}}is}, \citenamefont {Ka{\v{s}}alynas}, \citenamefont
  {Seniutinas}, \citenamefont {Juodkazis}, \citenamefont {Prystawko},\ and\
  \citenamefont {Leszczy{\'n}ski}}]{jakvstas17}%
  \BibitemOpen
  \bibfield  {author} {\bibinfo {author} {\bibfnamefont {V.}~\bibnamefont
  {Jak{\v{s}}tas}}, \bibinfo {author} {\bibfnamefont {I.}~\bibnamefont
  {Grigelionis}}, \bibinfo {author} {\bibfnamefont {V.}~\bibnamefont
  {Janonis}}, \bibinfo {author} {\bibfnamefont {G.}~\bibnamefont
  {Valu{\v{s}}is}}, \bibinfo {author} {\bibfnamefont {I.}~\bibnamefont
  {Ka{\v{s}}alynas}}, \bibinfo {author} {\bibfnamefont {G.}~\bibnamefont
  {Seniutinas}}, \bibinfo {author} {\bibfnamefont {S.}~\bibnamefont
  {Juodkazis}}, \bibinfo {author} {\bibfnamefont {P.}~\bibnamefont
  {Prystawko}},\ and\ \bibinfo {author} {\bibfnamefont {M.}~\bibnamefont
  {Leszczy{\'n}ski}},\ }\bibfield  {title} {\bibinfo {title} {Electrically
  driven terahertz radiation of 2deg plasmons in algan/gan structures at 110 k
  temperature},\ }\href@noop {} {\bibfield  {journal} {\bibinfo  {journal}
  {Appl. Phys. Lett.}\ }\textbf {\bibinfo {volume} {110}},\ \bibinfo {pages}
  {202101} (\bibinfo {year} {2017})}\BibitemShut {NoStop}%
\bibitem [{\citenamefont {Neto}\ \emph {et~al.}(2009)\citenamefont {Neto},
  \citenamefont {Guinea}, \citenamefont {Peres}, \citenamefont {Novoselov},\
  and\ \citenamefont {Geim}}]{neto09}%
  \BibitemOpen
  \bibfield  {author} {\bibinfo {author} {\bibfnamefont {A.~H.~C.}\
  \bibnamefont {Neto}}, \bibinfo {author} {\bibfnamefont {F.}~\bibnamefont
  {Guinea}}, \bibinfo {author} {\bibfnamefont {N.~M.~R.}\ \bibnamefont
  {Peres}}, \bibinfo {author} {\bibfnamefont {K.~S.}\ \bibnamefont
  {Novoselov}},\ and\ \bibinfo {author} {\bibfnamefont {A.~K.}\ \bibnamefont
  {Geim}},\ }\bibfield  {title} {\bibinfo {title} {The electronic properties of
  graphene},\ }\href@noop {} {\bibfield  {journal} {\bibinfo  {journal} {Rev.
  Mod. Phys.}\ }\textbf {\bibinfo {volume} {81}},\ \bibinfo {pages} {109}
  (\bibinfo {year} {2009})}\BibitemShut {NoStop}%
\bibitem [{\citenamefont {Grigorenko}\ \emph {et~al.}(2012)\citenamefont
  {Grigorenko}, \citenamefont {Polini},\ and\ \citenamefont
  {Novoselov}}]{grigorenko12}%
  \BibitemOpen
  \bibfield  {author} {\bibinfo {author} {\bibfnamefont {A.~N.}\ \bibnamefont
  {Grigorenko}}, \bibinfo {author} {\bibfnamefont {M.}~\bibnamefont {Polini}},\
  and\ \bibinfo {author} {\bibfnamefont {K.~S.}\ \bibnamefont {Novoselov}},\
  }\bibfield  {title} {\bibinfo {title} {Graphene plasmonics},\ }\href@noop {}
  {\bibfield  {journal} {\bibinfo  {journal} {Nat. Photon.}\ }\textbf {\bibinfo
  {volume} {6}},\ \bibinfo {pages} {749} (\bibinfo {year} {2012})}\BibitemShut
  {NoStop}%
\bibitem [{\citenamefont {Huang}\ \emph {et~al.}(2016)\citenamefont {Huang},
  \citenamefont {Song}, \citenamefont {Zhang},\ and\ \citenamefont
  {Yan}}]{huang17}%
  \BibitemOpen
  \bibfield  {author} {\bibinfo {author} {\bibfnamefont {S.}~\bibnamefont
  {Huang}}, \bibinfo {author} {\bibfnamefont {C.}~\bibnamefont {Song}},
  \bibinfo {author} {\bibfnamefont {G.}~\bibnamefont {Zhang}},\ and\ \bibinfo
  {author} {\bibfnamefont {H.}~\bibnamefont {Yan}},\ }\bibfield  {title}
  {\bibinfo {title} {Graphene plasmonics: physics and potential applications},\
  }\href@noop {} {\bibfield  {journal} {\bibinfo  {journal} {Nanophotonics}\
  }\textbf {\bibinfo {volume} {6}},\ \bibinfo {pages} {1191} (\bibinfo {year}
  {2016})}\BibitemShut {NoStop}%
\bibitem [{\citenamefont {Chen}\ \emph {et~al.}(2017)\citenamefont {Chen},
  \citenamefont {Argyropoulos}, \citenamefont {Farhat},\ and\ \citenamefont
  {Gomez-Diaz}}]{chen17}%
  \BibitemOpen
  \bibfield  {author} {\bibinfo {author} {\bibfnamefont {P.-Y.}\ \bibnamefont
  {Chen}}, \bibinfo {author} {\bibfnamefont {C.}~\bibnamefont {Argyropoulos}},
  \bibinfo {author} {\bibfnamefont {M.}~\bibnamefont {Farhat}},\ and\ \bibinfo
  {author} {\bibfnamefont {J.~S.}\ \bibnamefont {Gomez-Diaz}},\ }\bibfield
  {title} {\bibinfo {title} {Flatland plasmonics and nanophotonics based on
  graphene and beyond},\ }\href@noop {} {\bibfield  {journal} {\bibinfo
  {journal} {Nanophotonics}\ }\textbf {\bibinfo {volume} {6}},\ \bibinfo
  {pages} {1239} (\bibinfo {year} {2017})}\BibitemShut {NoStop}%
\bibitem [{\citenamefont {Ooi}\ and\ \citenamefont {Tan}(2017)}]{ooi17}%
  \BibitemOpen
  \bibfield  {author} {\bibinfo {author} {\bibfnamefont {K.~J.~A.}\
  \bibnamefont {Ooi}}\ and\ \bibinfo {author} {\bibfnamefont {D.~T.~H.}\
  \bibnamefont {Tan}},\ }\bibfield  {title} {\bibinfo {title} {Nonlinear
  graphene plasmonics},\ }\href@noop {} {\bibfield  {journal} {\bibinfo
  {journal} {Proc. R. Soc. A}\ }\textbf {\bibinfo {volume} {473}},\ \bibinfo
  {pages} {20170433} (\bibinfo {year} {2017})}\BibitemShut {NoStop}%
\bibitem [{\citenamefont {Guo}\ \emph {et~al.}(2017)\citenamefont {Guo},
  \citenamefont {Li}, \citenamefont {Deng}, \citenamefont {Yuan}, \citenamefont
  {Guinea},\ and\ \citenamefont {Xia}}]{guo17}%
  \BibitemOpen
  \bibfield  {author} {\bibinfo {author} {\bibfnamefont {Q.}~\bibnamefont
  {Guo}}, \bibinfo {author} {\bibfnamefont {C.}~\bibnamefont {Li}}, \bibinfo
  {author} {\bibfnamefont {B.}~\bibnamefont {Deng}}, \bibinfo {author}
  {\bibfnamefont {S.}~\bibnamefont {Yuan}}, \bibinfo {author} {\bibfnamefont
  {F.}~\bibnamefont {Guinea}},\ and\ \bibinfo {author} {\bibfnamefont
  {F.}~\bibnamefont {Xia}},\ }\bibfield  {title} {\bibinfo {title} {Infrared
  nanophotonics based on graphene plasmonics},\ }\href@noop {} {\bibfield
  {journal} {\bibinfo  {journal} {ACS Photonics}\ }\textbf {\bibinfo {volume}
  {4}},\ \bibinfo {pages} {2989} (\bibinfo {year} {2017})}\BibitemShut
  {NoStop}%
\bibitem [{\citenamefont {Fan}\ \emph {et~al.}(2019)\citenamefont {Fan},
  \citenamefont {Shen}, \citenamefont {Zhang}, \citenamefont {Zhao},
  \citenamefont {Wu}, \citenamefont {Fu}, \citenamefont {Wei}, \citenamefont
  {Li},\ and\ \citenamefont {Soukoulis}}]{fan19}%
  \BibitemOpen
  \bibfield  {author} {\bibinfo {author} {\bibfnamefont {Y.}~\bibnamefont
  {Fan}}, \bibinfo {author} {\bibfnamefont {N.~H.}\ \bibnamefont {Shen}},
  \bibinfo {author} {\bibfnamefont {F.}~\bibnamefont {Zhang}}, \bibinfo
  {author} {\bibfnamefont {Q.}~\bibnamefont {Zhao}}, \bibinfo {author}
  {\bibfnamefont {H.}~\bibnamefont {Wu}}, \bibinfo {author} {\bibfnamefont
  {Q.}~\bibnamefont {Fu}}, \bibinfo {author} {\bibfnamefont {Z.}~\bibnamefont
  {Wei}}, \bibinfo {author} {\bibfnamefont {H.}~\bibnamefont {Li}},\ and\
  \bibinfo {author} {\bibfnamefont {C.~M.}\ \bibnamefont {Soukoulis}},\
  }\bibfield  {title} {\bibinfo {title} {Graphene plasmonics: a platform for 2d
  optics},\ }\href@noop {} {\bibfield  {journal} {\bibinfo  {journal} {Adv.
  Optical Mater.}\ }\textbf {\bibinfo {volume} {7}},\ \bibinfo {pages}
  {1800537} (\bibinfo {year} {2019})}\BibitemShut {NoStop}%
\bibitem [{\citenamefont {Bandurin}\ \emph {et~al.}(2018)\citenamefont
  {Bandurin}, \citenamefont {Svintsov}, \citenamefont {Gayduchenko},
  \citenamefont {Xu}, \citenamefont {Principi}, \citenamefont {Moskotin},
  \citenamefont {Tretyakov}, \citenamefont {Yagodkin}, \citenamefont {Zhukov},
  \citenamefont {Taniguchi} \emph {et~al.}}]{bandurin18}%
  \BibitemOpen
  \bibfield  {author} {\bibinfo {author} {\bibfnamefont {D.~A.}\ \bibnamefont
  {Bandurin}}, \bibinfo {author} {\bibfnamefont {D.}~\bibnamefont {Svintsov}},
  \bibinfo {author} {\bibfnamefont {I.}~\bibnamefont {Gayduchenko}}, \bibinfo
  {author} {\bibfnamefont {S.~G.}\ \bibnamefont {Xu}}, \bibinfo {author}
  {\bibfnamefont {A.}~\bibnamefont {Principi}}, \bibinfo {author}
  {\bibfnamefont {M.}~\bibnamefont {Moskotin}}, \bibinfo {author}
  {\bibfnamefont {I.}~\bibnamefont {Tretyakov}}, \bibinfo {author}
  {\bibfnamefont {D.}~\bibnamefont {Yagodkin}}, \bibinfo {author}
  {\bibfnamefont {S.}~\bibnamefont {Zhukov}}, \bibinfo {author} {\bibfnamefont
  {T.}~\bibnamefont {Taniguchi}}, \emph {et~al.},\ }\bibfield  {title}
  {\bibinfo {title} {Resonant terahertz detection using graphene plasmons},\
  }\href@noop {} {\bibfield  {journal} {\bibinfo  {journal} {Nat. Commun.}\
  }\textbf {\bibinfo {volume} {9}},\ \bibinfo {pages} {1} (\bibinfo {year}
  {2018})}\BibitemShut {NoStop}%
\bibitem [{\citenamefont {Boubanga-Tombet}\ \emph {et~al.}(2020)\citenamefont
  {Boubanga-Tombet}, \citenamefont {Knap}, \citenamefont {Yadav}, \citenamefont
  {Satou}, \citenamefont {But}, \citenamefont {Popov}, \citenamefont
  {Gorbenko}, \citenamefont {Kachorovskii},\ and\ \citenamefont
  {Otsuji}}]{boubanga20}%
  \BibitemOpen
  \bibfield  {author} {\bibinfo {author} {\bibfnamefont {S.~A.}\ \bibnamefont
  {Boubanga-Tombet}}, \bibinfo {author} {\bibfnamefont {W.}~\bibnamefont
  {Knap}}, \bibinfo {author} {\bibfnamefont {D.}~\bibnamefont {Yadav}},
  \bibinfo {author} {\bibfnamefont {A.}~\bibnamefont {Satou}}, \bibinfo
  {author} {\bibfnamefont {D.~B.}\ \bibnamefont {But}}, \bibinfo {author}
  {\bibfnamefont {V.~V.}\ \bibnamefont {Popov}}, \bibinfo {author}
  {\bibfnamefont {I.~V.}\ \bibnamefont {Gorbenko}}, \bibinfo {author}
  {\bibfnamefont {V.~Y.}\ \bibnamefont {Kachorovskii}},\ and\ \bibinfo {author}
  {\bibfnamefont {T.}~\bibnamefont {Otsuji}},\ }\bibfield  {title} {\bibinfo
  {title} {Room-temperature amplification of terahertz radiation by
  grating-gate graphene structures},\ }\href@noop {} {\bibfield  {journal}
  {\bibinfo  {journal} {Phys. Rev. X}\ }\textbf {\bibinfo {volume} {10}},\
  \bibinfo {pages} {031004} (\bibinfo {year} {2020})}\BibitemShut {NoStop}%
\bibitem [{\citenamefont {Tomadin}\ and\ \citenamefont
  {Polini}(2013)}]{tomadin13}%
  \BibitemOpen
  \bibfield  {author} {\bibinfo {author} {\bibfnamefont {A.}~\bibnamefont
  {Tomadin}}\ and\ \bibinfo {author} {\bibfnamefont {M.}~\bibnamefont
  {Polini}},\ }\bibfield  {title} {\bibinfo {title} {Theory of the plasma-wave
  photoresponse of a gated graphene sheet},\ }\href@noop {} {\bibfield
  {journal} {\bibinfo  {journal} {Phys. Rev. B}\ }\textbf {\bibinfo {volume}
  {88}},\ \bibinfo {pages} {205426} (\bibinfo {year} {2013})}\BibitemShut
  {NoStop}%
\bibitem [{\citenamefont {Svintsov}\ \emph {et~al.}(2013)\citenamefont
  {Svintsov}, \citenamefont {Vyurkov}, \citenamefont {Ryzhii},\ and\
  \citenamefont {Otsuji}}]{svintsov13}%
  \BibitemOpen
  \bibfield  {author} {\bibinfo {author} {\bibfnamefont {D.}~\bibnamefont
  {Svintsov}}, \bibinfo {author} {\bibfnamefont {V.}~\bibnamefont {Vyurkov}},
  \bibinfo {author} {\bibfnamefont {V.}~\bibnamefont {Ryzhii}},\ and\ \bibinfo
  {author} {\bibfnamefont {T.}~\bibnamefont {Otsuji}},\ }\bibfield  {title}
  {\bibinfo {title} {Hydrodynamic electron transport and nonlinear waves in
  graphene},\ }\href@noop {} {\bibfield  {journal} {\bibinfo  {journal} {Phys.
  Rev. B}\ }\textbf {\bibinfo {volume} {88}},\ \bibinfo {pages} {245444}
  (\bibinfo {year} {2013})}\BibitemShut {NoStop}%
\bibitem [{\citenamefont {Mendl}\ and\ \citenamefont {Lucas}(2018)}]{mendl18}%
  \BibitemOpen
  \bibfield  {author} {\bibinfo {author} {\bibfnamefont {C.~B.}\ \bibnamefont
  {Mendl}}\ and\ \bibinfo {author} {\bibfnamefont {A.}~\bibnamefont {Lucas}},\
  }\bibfield  {title} {\bibinfo {title} {Dyakonov-shur instability across the
  ballistic-to-hydrodynamic crossover},\ }\href@noop {} {\bibfield  {journal}
  {\bibinfo  {journal} {Appl. Phys. Lett.}\ }\textbf {\bibinfo {volume}
  {112}},\ \bibinfo {pages} {124101} (\bibinfo {year} {2018})}\BibitemShut
  {NoStop}%
\bibitem [{\citenamefont {Mendl}\ \emph {et~al.}(2021)\citenamefont {Mendl},
  \citenamefont {Polini},\ and\ \citenamefont {Lucas}}]{mendl21}%
  \BibitemOpen
  \bibfield  {author} {\bibinfo {author} {\bibfnamefont {C.~B.}\ \bibnamefont
  {Mendl}}, \bibinfo {author} {\bibfnamefont {M.}~\bibnamefont {Polini}},\ and\
  \bibinfo {author} {\bibfnamefont {A.}~\bibnamefont {Lucas}},\ }\bibfield
  {title} {\bibinfo {title} {Coherent terahertz radiation from a nonlinear
  oscillator of viscous electrons},\ }\href@noop {} {\bibfield  {journal}
  {\bibinfo  {journal} {Appl. Phys. Lett.}\ }\textbf {\bibinfo {volume}
  {118}},\ \bibinfo {pages} {013105} (\bibinfo {year} {2021})}\BibitemShut
  {NoStop}%
\bibitem [{\citenamefont {Bistritzer}\ and\ \citenamefont
  {MacDonald}(2009)}]{bistritzer09}%
  \BibitemOpen
  \bibfield  {author} {\bibinfo {author} {\bibfnamefont {R.}~\bibnamefont
  {Bistritzer}}\ and\ \bibinfo {author} {\bibfnamefont {A.~H.}\ \bibnamefont
  {MacDonald}},\ }\bibfield  {title} {\bibinfo {title} {Hydrodynamic theory of
  transport in doped graphene},\ }\href@noop {} {\bibfield  {journal} {\bibinfo
   {journal} {Phys. Rev. B}\ }\textbf {\bibinfo {volume} {80}},\ \bibinfo
  {pages} {085109} (\bibinfo {year} {2009})}\BibitemShut {NoStop}%
\bibitem [{\citenamefont {Rudin}(2011)}]{rudin11}%
  \BibitemOpen
  \bibfield  {author} {\bibinfo {author} {\bibfnamefont {S.}~\bibnamefont
  {Rudin}},\ }\bibfield  {title} {\bibinfo {title} {Non-linear plasma
  oscillations in semiconductor and graphene channels and application to the
  detection of terahertz signals},\ }\href@noop {} {\bibfield  {journal}
  {\bibinfo  {journal} {Int. J. High Speed Electron. Syst.}\ }\textbf {\bibinfo
  {volume} {20}},\ \bibinfo {pages} {567} (\bibinfo {year} {2011})}\BibitemShut
  {NoStop}%
\bibitem [{\citenamefont {Svintsov}\ \emph {et~al.}(2012)\citenamefont
  {Svintsov}, \citenamefont {Vyurkov}, \citenamefont {Yurchenko}, \citenamefont
  {Otsuji},\ and\ \citenamefont {Ryzhii}}]{svintsov12}%
  \BibitemOpen
  \bibfield  {author} {\bibinfo {author} {\bibfnamefont {D.}~\bibnamefont
  {Svintsov}}, \bibinfo {author} {\bibfnamefont {V.}~\bibnamefont {Vyurkov}},
  \bibinfo {author} {\bibfnamefont {S.}~\bibnamefont {Yurchenko}}, \bibinfo
  {author} {\bibfnamefont {T.}~\bibnamefont {Otsuji}},\ and\ \bibinfo {author}
  {\bibfnamefont {V.}~\bibnamefont {Ryzhii}},\ }\bibfield  {title} {\bibinfo
  {title} {Hydrodynamic model for electron-hole plasma in graphene},\
  }\href@noop {} {\bibfield  {journal} {\bibinfo  {journal} {J. Appl. Phys.}\
  }\textbf {\bibinfo {volume} {111}},\ \bibinfo {pages} {083715} (\bibinfo
  {year} {2012})}\BibitemShut {NoStop}%
\bibitem [{\citenamefont {Briskot}\ \emph {et~al.}(2015)\citenamefont
  {Briskot}, \citenamefont {Sch{\"u}tt}, \citenamefont {Gornyi}, \citenamefont
  {Titov}, \citenamefont {Narozhny},\ and\ \citenamefont {Mirlin}}]{briskot15}%
  \BibitemOpen
  \bibfield  {author} {\bibinfo {author} {\bibfnamefont {U.}~\bibnamefont
  {Briskot}}, \bibinfo {author} {\bibfnamefont {M.}~\bibnamefont {Sch{\"u}tt}},
  \bibinfo {author} {\bibfnamefont {I.~V.}\ \bibnamefont {Gornyi}}, \bibinfo
  {author} {\bibfnamefont {M.}~\bibnamefont {Titov}}, \bibinfo {author}
  {\bibfnamefont {B.~N.}\ \bibnamefont {Narozhny}},\ and\ \bibinfo {author}
  {\bibfnamefont {A.~D.}\ \bibnamefont {Mirlin}},\ }\bibfield  {title}
  {\bibinfo {title} {Collision-dominated nonlinear hydrodynamics in graphene},\
  }\href@noop {} {\bibfield  {journal} {\bibinfo  {journal} {Phys. Rev. B}\
  }\textbf {\bibinfo {volume} {92}},\ \bibinfo {pages} {115426} (\bibinfo
  {year} {2015})}\BibitemShut {NoStop}%
\bibitem [{\citenamefont {Gantmakher}\ and\ \citenamefont
  {Levinson}(2012)}]{gantmakher12}%
  \BibitemOpen
  \bibfield  {author} {\bibinfo {author} {\bibfnamefont {V.~F.}\ \bibnamefont
  {Gantmakher}}\ and\ \bibinfo {author} {\bibfnamefont {Y.~B.}\ \bibnamefont
  {Levinson}},\ }\href@noop {} {\emph {\bibinfo {title} {Carrier scattering in
  metals and semiconductors}}}\ (\bibinfo  {publisher} {Elsevier},\ \bibinfo
  {year} {2012})\BibitemShut {NoStop}%
\bibitem [{\citenamefont {Dmitriev}\ \emph {et~al.}(1997)\citenamefont
  {Dmitriev}, \citenamefont {Furman}, \citenamefont {Kachorovskii},
  \citenamefont {Samsonidze},\ and\ \citenamefont {Samsonidze}}]{Dmitriev97}%
  \BibitemOpen
  \bibfield  {author} {\bibinfo {author} {\bibfnamefont {A.~P.}\ \bibnamefont
  {Dmitriev}}, \bibinfo {author} {\bibfnamefont {A.~S.}\ \bibnamefont
  {Furman}}, \bibinfo {author} {\bibfnamefont {V.~Y.}\ \bibnamefont
  {Kachorovskii}}, \bibinfo {author} {\bibfnamefont {G.~G.}\ \bibnamefont
  {Samsonidze}},\ and\ \bibinfo {author} {\bibfnamefont {G.~G.}\ \bibnamefont
  {Samsonidze}},\ }\bibfield  {title} {\bibinfo {title} {Numerical study of the
  current instability in a two-dimensional electron fluid},\ }\href
  {https://doi.org/10.1103/PhysRevB.55.10319} {\bibfield  {journal} {\bibinfo
  {journal} {Phys. Rev. B}\ }\textbf {\bibinfo {volume} {55}},\ \bibinfo
  {pages} {10319} (\bibinfo {year} {1997})}\BibitemShut {NoStop}%
\bibitem [{\citenamefont {Cheremisin}(2002)}]{cheremisin02}%
  \BibitemOpen
  \bibfield  {author} {\bibinfo {author} {\bibfnamefont {M.~V.}\ \bibnamefont
  {Cheremisin}},\ }\bibfield  {title} {\bibinfo {title} {Nonlinear regime of
  the current instability in a ballistic field effect transistor},\ }\href@noop
  {} {\bibfield  {journal} {\bibinfo  {journal} {Phys. Rev. B}\ }\textbf
  {\bibinfo {volume} {65}},\ \bibinfo {pages} {085301} (\bibinfo {year}
  {2002})}\BibitemShut {NoStop}%
\bibitem [{\citenamefont {Li}\ \emph {et~al.}(2017)\citenamefont {Li},
  \citenamefont {Hao}, \citenamefont {Jin},\ and\ \citenamefont {Lu}}]{li17}%
  \BibitemOpen
  \bibfield  {author} {\bibinfo {author} {\bibfnamefont {K.}~\bibnamefont
  {Li}}, \bibinfo {author} {\bibfnamefont {Y.}~\bibnamefont {Hao}}, \bibinfo
  {author} {\bibfnamefont {X.}~\bibnamefont {Jin}},\ and\ \bibinfo {author}
  {\bibfnamefont {W.}~\bibnamefont {Lu}},\ }\bibfield  {title} {\bibinfo
  {title} {Hydrodynamic electronic fluid instability in gaas mesfets at
  terahertz frequencies},\ }\href@noop {} {\bibfield  {journal} {\bibinfo
  {journal} {J. Phys. D: Appl. Phys.}\ }\textbf {\bibinfo {volume} {51}},\
  \bibinfo {pages} {035104} (\bibinfo {year} {2017})}\BibitemShut {NoStop}%
\bibitem [{\citenamefont {Bhardwaj}\ \emph {et~al.}(2016)\citenamefont
  {Bhardwaj}, \citenamefont {Nahar}, \citenamefont {Rajan},\ and\ \citenamefont
  {Volakis}}]{bhardwaj16}%
  \BibitemOpen
  \bibfield  {author} {\bibinfo {author} {\bibfnamefont {S.}~\bibnamefont
  {Bhardwaj}}, \bibinfo {author} {\bibfnamefont {N.~K.}\ \bibnamefont {Nahar}},
  \bibinfo {author} {\bibfnamefont {S.}~\bibnamefont {Rajan}},\ and\ \bibinfo
  {author} {\bibfnamefont {J.~L.}\ \bibnamefont {Volakis}},\ }\bibfield
  {title} {\bibinfo {title} {Numerical analysis of terahertz emissions from an
  ungated hemt using full-wave hydrodynamic model},\ }\href@noop {} {\bibfield
  {journal} {\bibinfo  {journal} {IEEE Trans. Electron Devices}\ }\textbf
  {\bibinfo {volume} {63}},\ \bibinfo {pages} {990} (\bibinfo {year}
  {2016})}\BibitemShut {NoStop}%
\bibitem [{\citenamefont {Nafari}\ \emph {et~al.}(2018)\citenamefont {Nafari},
  \citenamefont {Aizin},\ and\ \citenamefont {Jornet}}]{nafari18}%
  \BibitemOpen
  \bibfield  {author} {\bibinfo {author} {\bibfnamefont {M.}~\bibnamefont
  {Nafari}}, \bibinfo {author} {\bibfnamefont {G.~R.}\ \bibnamefont {Aizin}},\
  and\ \bibinfo {author} {\bibfnamefont {J.~M.}\ \bibnamefont {Jornet}},\
  }\bibfield  {title} {\bibinfo {title} {Plasmonic hemt terahertz transmitter
  based on the dyakonov-shur instability: Performance analysis and impact of
  nonideal boundaries},\ }\href@noop {} {\bibfield  {journal} {\bibinfo
  {journal} {Phys. Rev. Appl.}\ }\textbf {\bibinfo {volume} {10}},\ \bibinfo
  {pages} {064025} (\bibinfo {year} {2018})}\BibitemShut {NoStop}%
\bibitem [{\citenamefont {Taflove}\ and\ \citenamefont
  {Hagness}(2005)}]{taflove05}%
  \BibitemOpen
  \bibfield  {author} {\bibinfo {author} {\bibfnamefont {A.}~\bibnamefont
  {Taflove}}\ and\ \bibinfo {author} {\bibfnamefont {S.~C.}\ \bibnamefont
  {Hagness}},\ }\href@noop {} {\emph {\bibinfo {title} {Computational
  electrodynamics: the finite-difference time-domain method}}}\ (\bibinfo
  {publisher} {Artech house},\ \bibinfo {year} {2005})\BibitemShut {NoStop}%
\bibitem [{\citenamefont {Meric}\ \emph {et~al.}(2008)\citenamefont {Meric},
  \citenamefont {Han}, \citenamefont {Young}, \citenamefont {Ozyilmaz},
  \citenamefont {Kim},\ and\ \citenamefont {Shepard}}]{meric08}%
  \BibitemOpen
  \bibfield  {author} {\bibinfo {author} {\bibfnamefont {I.}~\bibnamefont
  {Meric}}, \bibinfo {author} {\bibfnamefont {M.~Y.}\ \bibnamefont {Han}},
  \bibinfo {author} {\bibfnamefont {A.~F.}\ \bibnamefont {Young}}, \bibinfo
  {author} {\bibfnamefont {B.}~\bibnamefont {Ozyilmaz}}, \bibinfo {author}
  {\bibfnamefont {P.}~\bibnamefont {Kim}},\ and\ \bibinfo {author}
  {\bibfnamefont {K.~L.}\ \bibnamefont {Shepard}},\ }\bibfield  {title}
  {\bibinfo {title} {Current saturation in zero-bandgap, top-gated graphene
  field-effect transistors},\ }\href@noop {} {\bibfield  {journal} {\bibinfo
  {journal} {Nat. Nanotech.}\ }\textbf {\bibinfo {volume} {3}},\ \bibinfo
  {pages} {654} (\bibinfo {year} {2008})}\BibitemShut {NoStop}%
\bibitem [{\citenamefont {Kapralov}\ and\ \citenamefont
  {Svintsov}(2020)}]{kapralov20}%
  \BibitemOpen
  \bibfield  {author} {\bibinfo {author} {\bibfnamefont {K.}~\bibnamefont
  {Kapralov}}\ and\ \bibinfo {author} {\bibfnamefont {D.}~\bibnamefont
  {Svintsov}},\ }\bibfield  {title} {\bibinfo {title} {Plasmon damping in
  electronically open systems},\ }\href@noop {} {\bibfield  {journal} {\bibinfo
   {journal} {Phys. Rev. Lett.}\ }\textbf {\bibinfo {volume} {125}},\ \bibinfo
  {pages} {236801} (\bibinfo {year} {2020})}\BibitemShut {NoStop}%
\bibitem [{\citenamefont {Crabb}\ \emph {et~al.}()\citenamefont {Crabb},
  \citenamefont {Cantos-Roman}, \citenamefont {Aizin},\ and\ \citenamefont
  {Jornet}}]{crabb21}%
  \BibitemOpen
  \bibfield  {author} {\bibinfo {author} {\bibfnamefont {J.}~\bibnamefont
  {Crabb}}, \bibinfo {author} {\bibfnamefont {X.}~\bibnamefont {Cantos-Roman}},
  \bibinfo {author} {\bibfnamefont {G.~R.}\ \bibnamefont {Aizin}},\ and\
  \bibinfo {author} {\bibfnamefont {J.~M.}\ \bibnamefont {Jornet}},\ }\bibinfo
  {title} {in preparation}\BibitemShut {NoStop}%
\bibitem [{\citenamefont {Dmitriev}\ \emph {et~al.}(2001)\citenamefont
  {Dmitriev}, \citenamefont {Kachorovskii},\ and\ \citenamefont
  {Shur}}]{dmitriev01}%
  \BibitemOpen
\bibfield  {title} {  }\bibfield  {author} {\bibinfo {author} {\bibfnamefont
  {A.~P.}\ \bibnamefont {Dmitriev}}, \bibinfo {author} {\bibfnamefont {V.~Y.}\
  \bibnamefont {Kachorovskii}},\ and\ \bibinfo {author} {\bibfnamefont {M.~S.}\
  \bibnamefont {Shur}},\ }\bibfield  {title} {\bibinfo {title} {Plasma wave
  instability in gated collisionless two-dimensional electron gas},\
  }\href@noop {} {\bibfield  {journal} {\bibinfo  {journal} {Appl. Phys.
  Lett.}\ }\textbf {\bibinfo {volume} {79}},\ \bibinfo {pages} {922} (\bibinfo
  {year} {2001})}\BibitemShut {NoStop}%
\bibitem [{\citenamefont {Cheremisin}\ and\ \citenamefont
  {Samsonidze}(1999)}]{cheremisin99}%
  \BibitemOpen
  \bibfield  {author} {\bibinfo {author} {\bibfnamefont {M.~V.}\ \bibnamefont
  {Cheremisin}}\ and\ \bibinfo {author} {\bibfnamefont {G.~G.}\ \bibnamefont
  {Samsonidze}},\ }\bibfield  {title} {\bibinfo {title} {D’yakonov-shur
  instability in a ballistic field-effect transistor with a spatially
  nonuniform channel},\ }\href@noop {} {\bibfield  {journal} {\bibinfo
  {journal} {Semiconductors}\ }\textbf {\bibinfo {volume} {33}},\ \bibinfo
  {pages} {578} (\bibinfo {year} {1999})}\BibitemShut {NoStop}%
\bibitem [{\citenamefont {Svintsov}(2018)}]{svintsov18}%
  \BibitemOpen
  \bibfield  {author} {\bibinfo {author} {\bibfnamefont {D.}~\bibnamefont
  {Svintsov}},\ }\bibfield  {title} {\bibinfo {title} {Exact solution for
  driven oscillations in plasmonic field-effect transistors},\ }\href@noop {}
  {\bibfield  {journal} {\bibinfo  {journal} {Phys. Rev. Appl.}\ }\textbf
  {\bibinfo {volume} {10}},\ \bibinfo {pages} {024037} (\bibinfo {year}
  {2018})}\BibitemShut {NoStop}%
\end{thebibliography}%
\end{document}